\RequirePackage{fix-cm}
\documentclass[smallextended]{svjour3}
\smartqed
\usepackage{graphicx}
\usepackage{amsmath}
\usepackage{bm}
\usepackage{subfig}
\usepackage{psfrag}

\def\bra#1{\mathinner{\langle{#1}|}}
\def\ket#1{\mathinner{|{#1}\rangle}}
\def\braket#1{\mathinner{\langle{#1}\rangle}}

{\catcode`\|=\active\gdef\Braket#1{\left<\mathcode`\|"8000\let|\bravert {#1}\right>}}
\def\bravert{\egroup\,\vrule\,\bgroup}
\def\sen{\mathop{\mbox{\normalfont sen}}\nolimits}
\def\sen{\mathop{\mbox{\normalfont sen}}\nolimits}

\usepackage[colorlinks=true, linkcolor=blue, citecolor=blue, urlcolor=black]{hyperref}
\AtBeginDocument{%
	\setlength{\oddsidemargin}{\dimexpr(\paperwidth-\textwidth)/2-1in}%
	\setlength{\evensidemargin}{\oddsidemargin}%
	\setlength{\topmargin}{%
		\dimexpr(\paperheight-\textheight)/2-\headheight-\headsep-1in}%
}

\usepackage{floatrow}

\begin{document}


\title{Relationship between the field local quadrature and the quantum discord  of a photon-added correlated channel under the influence of scattering and phase fluctuation noise}
\author{Francisco A. Dom\'inguez-Serna  \and Francisco J. Mendieta-Jimenez \and
       Fernando Rojas 
}
\institute{F. A. Domínguez Serna \at
              Posgrado en F\'isica de Materiales, Centro de Investigaci\'on Cient\'ifica y de Educaci\'on Superior de Ensenada, Ensenada 22890, Baja California, M\'exico.
              \email{francisco.ds@gmail.com}           
			\and F. J. Mendieta-Jimenez \at
				Agencia Espacial Mexicana, Xola y Universidad, D.F., 03020, M\'exico\\
				\email{mendieta.javier@aem.gob.mx}
           \and
           F. Rojas \at
              	Centro de Nanociencias y Nanotecnolog\'ia – Departamento de F\'isica Te\'orica,  Universidad Nacional Aut\'onoma de M\'exico, Ensenada, Baja California, 22860 M\'exico\\
              	\email{frojas@cnyn.unam.mx}
}
\date{	Published in 2017. ``The final publication is available at Springer via \\ http://dx.doi.org/10.1007/s11128-017-1704-x".	}

\maketitle

\begin{abstract}
We study quantum correlations  and discord in a bipartite continuous variable hybrid system formed by linear combinations of coherent states $\ket{\alpha}$ and single photon added coherent states (SPACS) of the form $\ket{\psi}_{\text{dp(pa)}}= \mathcal{N}/\sqrt{2} (\hat{a}^\dagger \ket{\alpha}_a\ket{\alpha}_b \pm \hat{b}^\dagger \ket{\alpha}_a\ket{\alpha}_b)$. We stablish a relationship between the quantum discord with a local observable (the quadrature variance for one sub-system) under the influence of scattering and phase fluctuation noise. For the pure states the quantum correlations are characterized by means of measurement induced disturbance (MID) with simultaneous quadrature measurements. In a scenario where homodyne conditional measurements are available we show that the  MID provides an easy way to select optimal phases to obtain information of the maximal correlations in the channels. The quantum correlations of these entangled states with channel losses  are quantitatively characterized with the quantum discord (QD) with a displaced qubit projector. We observe that as scattering increases, QD decreases monotonically. At the same time for the state $\ket{\psi}_{\text{dp}}$, QD is more resistant to high phase fluctuations when the average photon number $n_0$ is bigger than zero, but if phase fluctuations are low, QD is more resistant if $n_0=0$. For the dp model with scattering, we obtain an analytical expression  of the QD as a function of the observable quadrature variance in a local sub-system. This relation  allows us to have a way to obtain the degree of QD in the channel  by just measuring a local property observable such as the quadrature variance. For the other model this relation still exists but is explored numerically. This relation is an important result that allows to identify quantum processing capabilities in terms of just local observables.

\end{abstract}


\section{\label{sec:level1}Introduction\\}
Quantum entanglement as a measure of non-separability of multipartite quantum states and its dynamics have been widely studied, both theoretically and experimentally, and is identified as a key resource for quantum information processing (QIP) \cite{Aolita2015,Jozsa2003}. Most applications of entanglement have been proposed in the discrete variable (DV) regime, in which they require highly precise detectors and demanding experiments \cite{Driessen2013}. Despite these difficulties, the entangled systems involving different degrees of freedom deserve special mention. For example, in the works of Mirza, and Mirza and Schotland \cite{Mirza2016,Mirza2015,Mirza2016a}, the dynamics of entanglement generation for atom-cavity arrangements are studied, the authors demonstrate that by mediating the coupling strength between systems, number of interacting atoms and excitation decay rates, entanglement among the different degrees of freedom of the involved systems can be controlled. Systems involving photonic subsystems and other parties of different nature entangled are of great importance, since these kind of states could be used to exploit QIP capabilities of different technologies with an easy integration in the already existent optical fiber network \cite{Obrien2010}. Experimental implementations of light-matter entangled states have already been made, for example entangling photons with solid-state qubits \cite{Togan2010}, single photons with single trapped atoms \cite{Volz2006}, polarization states of single photons with a single ion \cite{Stute2013}, and many other implementations that can be found in the references therein. On the other hand, continuous variable (CV) QIP has been recently explored \cite{Braunstein2005}, which at least makes detection experiments easier with the cost that the experimental complexity to generate those CV entangled states increases in comparison with DV set ups \cite{Masada15}. To exploit the advantages of CV-QIP research in the hybrid CV-DV regime has also been made, as shown, for instance, in  \cite{Takeda2013a} where a CV channel is used to encode and teleport a discrete variable, and \cite{Sherson2006} where a light pulse is teleported through an entangled atomic ensemble. All these entangled states with systems of different degrees of freedom have potential applications in quantum communications protocols (QCP), and in general in QIP.

Even though entangled states are necessarily non-separable, there exists some separable states that exhibit non classical correlations \cite{Gu2012,Modi2012,Laflamme2001} and have proven usefulness as resources for QIP and quantum communications. Quantum discord (QD) is a known measure to quantify these correlations with no classical counterpart. QD quantifies the difference between mutual quantum information and classical correlations (CC) \cite{Modi2012}. Mutual information measures the amount of knowledge that can be gained by measuring one of the parties regarding the other in a bipartite system, i.e., the non classical correlations between them. CC are generated by the measurement projection on one subsystem and its effect on the other. QD is a proven resource to remote state preparation, with more importance than entanglement \cite{Dakic2012a}, and has also been studied to quantify the 'quantum advantage' of one state over some others \cite{Gu2012} when the state is used in a QCP. In a similar context, has been shown that for multiparty systems, the solely interaction with the environment can generate QD, i.e., the noise can create non-classical correlations \cite{Lanyon2013}. The experimental characterization of QD in DV systems can be overcome with the use of density matrix reconstruction techniques and further evaluation of QD, which has been applied for instance, to estimate QD for polarization qubits obtained by a non-linear crystal under the influence of noise \cite{Benedetti2013}. The time dynamics of QD differs from the ones of quantum entanglement, and result of great interest since all systems are immersed in a decoherent medium \cite{Fanchini2010,Modi2012}. In this respect, we can find recent studies about experimental dynamics of QD in optical, solid-state spin and nuclear magnetic resonance systems \cite{FernandesFanchini2017}. Particularly for optical systems, where interactions are well controlled, the dynamical behavior of QD has been used to study transitions between classical and quantum correlations  \cite{Xu2017}. In the CV regime has been proven that quadrature measurements could be used as a tool for experimental verification of QD in Gaussian and certain non-Gaussian states \cite{Hosseini2014,Hosseini2014a}. In this paper we study non-classical indicators for bipartite systems composed by CV and DV parties. Given the hybrid structure of the states under study with loss mechanisms of scattering and phase fluctuations  the  correlations are better characterized with a quantitative QD based indicators instead of the quantum entanglement alone \cite{Ferraro2010}.

A coherent state (CS) is a CV state that can be easily generated by a laser and its classical properties are well known. Universal quantum computing with CS was proposed by Jeong et. al. \cite{Jeong08a} with potential applications in communications. But, it relies on coherent entangled states, which are difficult to generate experimentally \cite{Daoud2012}. On the contrary, it is possible to obtain a hybrid entangled state (HES) formed by a linear combination of CS and another non-classical state. Fock states $\ket{n}$ as purely quantum, are considered the most non-classical states. A single photon added coherent state (SPACS) is generated by the single application of the creation operator over a coherent state $\propto \hat{a}^\dagger \ket{\alpha}$, and has combined properties of a classical and a non-classical state. SPACS where first proposed by Agarwal and Tara \cite{Agarwal1991} and where experimentally obtained later by the interaction of a CS with a non-linear crystal \cite{Zavatta2004}, and also are useful to study the transition between classical and quantum domains. SPACS are also characterized by a negativity in the Wigner function, which is a signature of non-classicality \cite{Kenfack2004}. Therefore, its use to generate bipartite entangled states through linear optics as a beamsplitter is also feasible \cite{Kim2002}.

An example of HES was proposed by \cite{Sekatski2012a}, where they combined CS with Fock states in a beamsplitter producing a superposition of displaced Fock states (DFS) and CS obtaining the wavefunction $\ket{\psi}=2^{-1/2}(\hat{D}(\alpha) \ket{1}\ket{\alpha} - \ket{1}\hat{D}(\alpha)\ket{1})$, which is equivalent to a superposition of SPACS with CS in the form $\ket{\psi}=2^{-1/2}( \hat{a}^\dagger \ket{\alpha}\ket{\alpha} - \hat{b}^\dagger \ket{\alpha}\ket{\alpha})$. We name this superposition as displaced fock (dp) state in the text. It is this implication what gives origin to the study of states of the form $\ket{\psi}_{\text{pa(pd)}}= \hat{a}^\dagger \ket{\alpha}\ket{\alpha} \pm \hat{b}^\dagger \ket{\alpha}\ket{\alpha}$ (up to a normalization constant), with correlation properties oriented to quantum communication applications. We name the positive superposition as photon added (pa) state, as it can be generated with a SPACS and a beamsplitter (BS). The states under study can be easily produced experimentally by means of linear operations with a nonlinear process required only for the photon addition part. Photon addition is an interesting process itself as it is part of non-Gaussian operations applied in quantum information with continuous variables \cite{Wang2015}. Photon addition operations have been also used in a systematic method to generate entanglement between subsystems of CS and single Fock states, as proposed and experimentally demonstrated by \cite{Jeong2014,Kwon2015}.

The superposition of SPACS and CS is of special interest given their hybrid nature between classical and quantum systems as CS represent a completely classical system, and the SPACS can act as a classical state when the average photon number is big enough, or quantum if the average photon number is low \cite{Zavatta2004,Kenfack2004}. The entanglement between these two degrees of freedom enables the proposition of these states as a suitable option to be explored for QIP in an also hybrid CV-DV regime. This is also reinforced with the experimental demonstration of SPACS, and its use in studying transitions between classical and quantum states.

A possible application of this bipartite correlated channels is in quantum communications. Quantum communications lie on the establishment of a protocol that requires the selection of a set of distinguishable measurements under a suitable strategy. Quadrature measurements result as a first option to consider when compared to other communication protocols \cite{Silva2006,Chuan2010,Paris2003}. Quadrature measurements are directly obtained by means of homodyne detections (HD) methods, which are standard in quantum communication protocols as well.

A first approach for a pure bipartite system, that one can use is the Joint Quadrature Probability (JQP) distribution, which is obtained by a simultaneous measurement of the quadrature of the bipartite state and its statistics. The use of JQP is initially proposed for its simplicity, and even though it can not show directly the amount of quantum correlations, it is useful in the sense that it indicates the existence of some correlations in the bipartite state used. Later, if some degree of correlation is found, other techniques can be used to remove the classical contribution and to obtain a quantitative characterization. By using HD projectors of the field, correlations can be characterized by ameliorated measurement induced disturbance (AMID) \cite{Ye2013,Girolami2011}. AMID is the difference between mutual quantum information (MQI) and classical correlations (CC) when both subsystems are simultaneously measured and then it is minimized through all possible measurements. It can be understood as the maximal entropic cost of any possible measurement given a complete set of measurements. If the optimization process is disregarded, then a more relaxed form called measurement induced disturbance (MID) is obtained \cite{Modi2010}. It also plays an important role when defining behavior of correlations given a projective measure.

It is important to note that the mechanisms to obtain non-classical features described above are only used in the pure channel because they are proposed as a semi quantitative tool  to characterize states being prepared in a laboratory, i.e., as the experiment is set, the bipartite generated state must be evaluated prior to its utilization as a quantum channel. At this stage, the bipartite state is ideally not subjected to noise, or at least to minimum sources of noise, as it presumably remains in a controlled environment. In this paper we study two different superpositions of bipartite SPACS, for which JQP and AMID are good candidates to partially characterize its correlations properties, and to differentiate  one from the other. Once the bipartite correlated state is generated, then it could be used as a quantum channel and resource for QIP. At this point, we introduce two common effects of noise in the channel, which are scattering and phase fluctuations, both of which affect the classical and quantum correlations of the channel, and therefore its QD. In this scenario we evaluate the  QD  in order to assess quantitatively the quantum properties of the channel. However, QD measures are not simple to do and we ask ourselves whether it is possible to use an easily accessible observable to determine the amount of QD of the channel when it is subjected to noise sources. It is found that the quadrature fluctuation $(\Delta X)^2$ of a subsystem, which can be measured in any experiment, could  be a good estimator of QD for these channels. Therefore we plan to  compare QD with this observable for the proposed channels to evaluate its functional relation.

In summary, the aim of this study is to quantify the non-classical signatures of bipartite superpositions of SPACS, that could be used as a channel intended to establish quantum communication protocols. Specially, we study how the channel is affected by loss mechanisms and the impact of these mechanisms on the quantum correlations of the bipartite state. Our work can be  divided in two main parts: the generation of the bipartite states formed by two different superpositions of SPACS with its qualitative characterization and the later quantification of QD with an easily accessible observable as described above.

This paper is organized as follows. In the introduction, brief descriptions of the non-classicality measures such as simultaneous homodyne detections, QD and MID are addressed. In section two, the model under study is fully described, first in a general scheme to create the states and second, the bipartite channel in a communication scenario is outlined, where one of the parties is kept local by Bob and the other one is sent to a distant place where Alice is located, subjected to loss mechanisms of scattering and phase noise. Section three includes a discussion of the obtained results of the correlations study and Alice's information gain given possible manipulations performed by Bob, whereby, QD from B to A is characterized, and its parametrized version to quadrature variance is estimated. Finally, conclusions of the study and comments for future work  are included.

\subsection{Quadrature conditional probabilities}

HD consist on mixing a field signal to be measured, generally of low intensity, with a strong signal of the same frequency known as the local oscillator (LO) by means of a BS whose output photocurrent are subtracted and proportional to the quadrature operator \cite{Collett1987,Barnett2002}

\begin{equation}
	\hat{X}_{\lambda_{A(B)}}=\frac{1}{\sqrt{2}}(\hat{a}(\hat{b})e^{-i\lambda_{A(B)}}+\hat{a}^\dagger (\hat{b}^\dagger)e^{i\lambda_{A(B)}}),
	\label{eq:QuadratureOp}
\end{equation}
where $\{\hat{a}^\dagger, \hat{a}\} (\{\hat{b}^\dagger, \hat{b}\})$ are photon creation and destruction operator for the corresponding subsystem $A(B)$, at the phase $\lambda_{A(B)}$ of the LO.

A simple communication strategy between two parties (Alice and Bob) sharing a common resource described by the density matrix $\rho^{AB}$ could rely on the correlations of the outcomes for the simultaneous positive operator-valued measure (POVM) of HD. Within each subsystem the set of POVM elements are designated as
\begin{equation}
\hat{\Pi}^{A(B)}(\lambda_{A(B)})=M^{A(B)}_{\lambda_{A(B)}} M^{A(B)\dagger}_{\lambda_{A(B)}}=\ket{X_{\lambda_{A(B)}}}\bra{X_{\lambda_{A(B)}}},
\label{eq:povm_hom}
\end{equation}
where the quadratures $\ket{X_{\lambda_{A(B)}}}$  are eigenstates of the operator (\ref{eq:QuadratureOp}) \cite{Barnett2002} and are given by $\ket{X_{\lambda_{A(B)}}} =\pi^{-\frac{1}{4}} \exp \{ -\frac{1}{2}X_{\lambda_{A(B)}}^2+\sqrt{2} e^{i\lambda}X_{\lambda_{A(B)}} \hat{a}^\dagger (\hat{b}^\dagger) -\frac{1}{2}e^{2i\lambda_{A(B)}}\hat{a}^{\dagger^2} (\hat{b}^{\dagger^2}) \}\ket{0}$.

The measurement operations are described as follows: Alice measures with her quadrature operator $\hat{\Pi}^A(\lambda_A)$ in the shared $ \rho^{AB} $ and obtains the outcome for $X_{\lambda_A}$ with probability $\text{P}_{A}(X_{\lambda_A})=\text{Tr}\{ \hat{\Pi}^{A} \otimes \mathbf{1}^B  \rho^{AB} \}$, what projects the state to $\rho^{AB} \rightarrow \rho^{AB}_{\Pi^A}$. Next, Bob measures with his quadrature operator $\hat{\Pi}^B(\lambda_B)$ and obtains $X_{\lambda_B}$ with the conditional probability $\text{P}_{B}(X_{\lambda_B}|X_{\lambda_A})=\text{Tr}[\mathbf{1}^A  \otimes \hat{\Pi}^{B} \rho^{AB}_{\Pi^A} ]$. And the joint probability of obtaining both $X_{\lambda_A}$ and $X_{\lambda_B}$ is obtained as

\begin{equation}
\begin{split}
\text{P}_{AB}(X_{\lambda_A},X_{\lambda_B})&=\text{P}_{A}(X_{\lambda_A})\text{P}_{B}(X_{\lambda_B}|X_{\lambda_A})\\
&=\text{Tr} [\hat{\Pi}^{A} \otimes \hat{\Pi}^{B} \rho^{AB}  ].
\end{split}
\label{eq:jointprobquad}
\end{equation}
Further knowledge on conditional and simulateneous measurements can be gained in QIP textbooks like the one of Audretsch \cite{Audretsch}. By using this simple JQP distribution we want to gain knowledge of the no classical behavior of the pure channel under study by means of HD that can be useful for experimental verification. This is so, because one can describe the density matrix of the system in the homodyne basis (that will be described below), which captures the continuous nature of the optical fields. This also, facilitates the expression of the projected system as a statistical mixture of the simultaneous HD outcomes that will be useful for homodyne MID calculations.  This process is also useful to select specific LO phases${\lambda_ A ,\lambda_B}$ that maximize the obtainable correlations between the two parties of the system.

\subsection{Measurement induced disturbance}
The measurement induced disturbance (MID) in quantum systems can be used to quantify their quantum correlations as proposed by Luo \cite{Luo2008}, which is based on the idea that measurements do not disturb a classical system, but only one with some degree of inherent quantumness. We start with the mutual information $\mathcal{I}(\rho^{AB})$ that accounts for all kind of correlations in a given system $\rho^{AB}$ and is determined as
\begin{equation}
\mathcal{I}(\rho^{AB})=S(\rho^A)+S(\rho^B)-S(\rho^{AB}),
\label{eq:MutInf}
\end{equation}
where $S$ is the Von Neumann entropy in the form $S(\rho)=-\text{Tr}[\rho \log \rho]$ \cite{Nielsen00a}, which we name entropy alone without distinction in the rest of the paper.

For bipartite systems, when $I(\rho^{AB})>0$,  measuring a subsystem alone can give information on the other.  The density matrix of a given system after a non-selective measurement can be expanded as  $\rho_{\Pi}^{AB}=\sum_i\sum_j \hat{\Pi}_i^A \otimes \hat{\Pi}_j^B \rho^{AB} \hat{\Pi}_i^A \otimes \hat{\Pi}_j^B =\sum_i\sum_j  p_{ij} \Pi_i^A \otimes \Pi_j^B$, where $\hat{\Pi}^{A(B)}_i$ form a complete set of POVM elements of the measurement, and $p_{ij} = \text{ Tr} [\Pi_i^A \otimes \Pi_j^B \rho^{AB} ]$. $\rho^{AB}_{\Pi}$ is a classical state in the sense that it is described as a statistical mixture of the possible outcomes of a set of measurements in both subsystems with its correspondent probability. Therefore its mutual information  $\mathcal{I}(\rho^{AB}_{\Pi})$ will give the classical correlations for that POVMs. The difference between mutual informations of the system before and after measurement will give the MID expressed as
\begin{equation}
D_{\text{MID}}(\rho^{AB})= \mathcal{I}(\rho^{AB})- \mathcal{I}(\rho^{AB}_{\Pi}).
\label{eq:MIDdef}
\end{equation}
MID will be zero only if the system can initially be expressed into the spectral decomposition of the POVMs used, i.e., if it only has classical correlations.

The optimized version of the MID, the Ameliorated MID (AMID) is:
\begin{equation}
\mathcal{A}(\rho^{AB})=\underset{\{\Pi_i\}}{\text{inf}} [ \mathcal{I}(\rho^{AB})- \mathcal{I}(\rho^{AB}_{\Pi}) ],
\label{eq:AMIDdef}
\end{equation}
where the minimum is taken over all possible measurements.

In this proposal, we are interested in studying the effects on non-classicality by homodyne measurements. If a non selective quadrature measurement is made on a subsystem (of a bipartite system), it drives the other one into a statistical mixture of infinite possible quadratures related to each possible phase angle of the local oscillator (LO). If both subsystems quadratures are simultaneously measured the full system is a statistical mixture of possible eigenstates of quadrature. A plausible method to study a channel with these characteristics is a quadrature MID approach, with a complete set of quadrature eigenprojectors, as they can be implemented by homodyne detections easily on practice \cite{Olivares2013,Kumar2012}. In addition, quadrature MID will also show how to select specific phase angles $\lambda_{A(B)}$ of the LO that maximize the classical  correlations between Alice and Bob for HD and characterize the non-classical correlations.

To apply the MID formalism in the quadrature setting, the full density matrix of the system is initially expanded in the quadrature eigenbasis as
\begin{equation}
\begin{split}
\rho^{AB}&=\int_{-\infty}^\infty dX'_{\lambda_A}dX'_{\lambda_B}  dX''_{\lambda_A}dX''_{\lambda_B} \ket{X'_{\lambda_A}}\bra{X'_{\lambda_A}}\\ & \otimes \ket{X'_{\lambda_B}}\bra{X'_{\lambda_B}} \rho^{AB} \ket{X''_{\lambda_A}}\bra{X''_{\lambda_A}}\otimes\ket{X''_{\lambda_B}}\bra{X''_{\lambda_B}}.
\end{split}
\label{eq:MID3}
\end{equation}
MID considers simultaneous measurement, therefore, the density matrix of the quadrature non-selective simultaneous measured system is
\begin{equation}
\begin{split}
\rho^{AB}_\Pi&=\int_{-\infty}^\infty dX_{\lambda_A}dX_{\lambda_B} P_{AB}(X_{\lambda_A}(\lambda_A),X_{\lambda_B}(\lambda_B)) \\ &\quad \times \ket{X_{\lambda_A}}\bra{X_{\lambda_A}}\otimes \ket{X_{\lambda_B}}\bra{X_{\lambda_B}},
\end{split}
\label{eq:MID2}
\end{equation}
where, $P_{AB}(X_{\lambda_A}(\lambda_A),X_{\lambda_B}(\lambda_B)) $ is the joint probability of the detection of a pair of quadratures $\{X_{\lambda_A},X_{\lambda_B} \}$ as in (\ref{eq:jointprobquad}), for a given $\lambda_A$ and $\lambda_B$. The notation of $P_{AB}$ has been used to emphasize the dependency with $\lambda_A$ and $\lambda_B$. Integrating over all the possible outcomes, the projected state is obtained. The mutual information of the system before measure $\mathcal{I}(\rho^{AB})$ is simply twice the entropy of entanglement, for this particular case considered as a pure system. This part is carried out in the original basis as it is simpler than in the quadrature basis. Also to obtain (\ref{eq:MIDdef}) and (\ref{eq:AMIDdef}), entropies of the projected states $S(\rho^{A}_\Pi)$, $S(\rho^{B}_\Pi)$ and $S(\rho^{AB}_\Pi)$ have to be calculated. These entropies are obtained with the quadrature expansion. The joint projected density matrix is diagonal in this basis, which allows an easy calculation of the entropy as
\begin{equation}
\begin{split}
S(\rho^{AB}_\Pi)&=\int_{-\infty}^\infty dX_{\lambda_A}dX_{\lambda_B} P_{AB}(X_{\lambda_A}(\lambda_A),X_{\lambda_B}(\lambda_B))\\ &\quad \times \log P_{AB}(X_{\lambda_A}(\lambda_A),X_{\lambda_B}(\lambda_B)).
\end{split}
\label{eq:MID6}
\end{equation}
For the remaining terms it is reduced to,
\begin{equation}
\begin{split}
S(\rho^{A(B)}_\Pi)&=\int_{-\infty}^\infty dX_{\lambda_{A(B)}} P_{A(B)}(X_{\lambda_{A(B)}}(\lambda_{A(B)}))\\ &\quad \times \log P_{A(B)}(X_{\lambda_{A(B)}}(\lambda_{A(B)})),
\end{split}
\label{eq:MID5}
\end{equation}
where the partial traces are changed for marginal integrals. The mutual information for the projected system  can now be obtained as
\begin{equation}
\mathcal{I}(\rho^{AB}_{\Pi}) ]=S(\rho^{A}_\Pi)+S(\rho^{B}_\Pi)-S(\rho^{AB}_\Pi).
\label{eq:I_proy}
\end{equation}

An important point for this quadrature MID and AMID approach is that in this work it is only used with pure states, as our intention at this stage is just to know whether there exist quantum correlations in the states by means of a JQP distribution that could be acquired in an experiment. A more exact quantification of these correlations when loss mechanisms are involved will be carried out by means of quantum discord in the following section.

\subsection{Quantum discord}
When a bipartite state is used to establish a realistic quantum communication channel, it undergoes through different types of loss. The different loss mechanisms induced by the medium in which the channel travels generate a statistical mixture related to the loss degrees of freedom and all the possible ways in which the loss could occur. Unlike pure states, entropy of entanglement is not adequate to characterize the non-classical behavior of the mixed states. In this context, QD quantitatively characterizes the quantum correlations even in the presence of mixing processes.

The idea of non-classicality as a measure related to mutual quantum information started with the seminal works of Ollivier and Zurek \cite{Ollivier2001} and Henderson and Vedral \cite{Henderson2001} and among the different proposals to quantify this non-clasicality, QD is the most used \cite{Giorda2012}.  Lets assume subsystem B, with reduded density matrix $\rho^B$ is measured with a positive operator-valued measure (POVM) and we want to gain information on the subsystem A, with reduced density matrix $\rho^A$. $\rho^{A(B)}$ is the reduced density matrix  $\rho^{A(B)}=\text{Tr}_{A(B)} [ \rho^{AB} ]$. The entropy of the full system $S(\rho^{AB})$ corresponds to the amount of information of the system  before any measurement. The POVM acting on $\rho^{B}$ has positive elements $\{ \hat{\Pi}_i^B=M_i^BM_i^{B\dagger}   \}$ that form a complete set $ \sum_i \hat{\Pi}_i^B=\mathbf{1}  $.  Each measurement on B projects the full density matrix in the form $\rho^{AB}\rightarrow \rho^{AB}_{ \Pi^B_i}=  \hat{\Pi}_i^B \rho^{AB} \hat{\Pi}_i^B / \text{Tr} [\rho^{AB} \Pi_i^B ]$, where the outcome of each measurement has a probability $p_i=\text{Tr} [\rho^{AB} \Pi_i^B ]$. The state of A after measurement on B is simply $\rho^{A}_{ \Pi^B_i}=\text{Tr}_B [\rho^{AB}_{ \Pi^B_i}]$. The entropy of A after measurement on B is obtained as $S(\rho^A_{\Pi^B})(A|B)= \sum_i p_i S(\rho^{A}_{ \Pi^B_i})$, here $\rho^A_{\Pi^B}$ stands for the ensemble $\rho^A_{\Pi^B}=\sum_i p_i \rho_{ \Pi^B_i}^{A}$ resulting from a non-selective measure on subsystem $B$ with its corresponding POVM elements, which is stated with the argument $(A|B)$.  Given that the amount of information of a subsystem is related to its entropy, the difference $S(\rho^A)-S(\rho^A_{\Pi^B})$ is related to the amount of information gained by the measurement process, and the maximal of this difference is the classical correlation present in the system,
\begin{equation}
\begin{split}
J_{cl}(A|B)&=\underset{\{\Pi_i\}}{\text{max}} [S(\rho^A)-S(\rho^A_{\Pi^B})(A|B)],\\
& = S(\rho^A) -  \underset{\{\Pi_i\}}{\text{min}} [S(\rho^A_{\Pi^B})(A|B)].
\end{split}
\end{equation}
Therefore with the aid of (\ref{eq:MutInf}), QD is defined as the subtraction of the mutual information (all kinds of correlations) from the full classical accessible correlation and is given as
\begin{equation}
\begin{split}
D_B(\rho^{AB})&=\mathcal{I}(\rho^{AB})-J_{cl}(A|B),\\
& = S(\rho^B)-S(\rho^{AB})+\underset{\{\Pi_i\}}{\text{min}} [S(\rho^A_{\Pi^B})(A|B)],
\end{split}
\label{eq:QD}
\end{equation}
where $D_B(\rho^{AB})$ is the QD when measurements are carried on $B$ subsystem. QD can be seen as measurement of the quantum correlation present in the state $\rho^{AB}$. QD is the property that we use to characterize the non-classical correlations of the channel when noise effects are included quantitatively.

\section{Model}

We are interested in the family of the photon added entangled states of the form
\begin{equation}
	\ket{\psi}=\frac{\mathcal{N}}{\sqrt{2}}(\hat{a}^\dagger \ket{\alpha}_a\ket{\alpha}_b \pm \hat{b}^\dagger \ket{\alpha}_a\ket{\alpha}_b),
	\label{eq:familyofstates}
\end{equation}
where $\mathcal{N}$ is a normalization constant, the subscripts represent a subsystem, with the photon creation operators $\hat{a}^\dagger$ and $\hat{b}^\dagger$ acting on the corresponding subsystem. We assume the equivalences $\ket{\alpha}_a\ket{\alpha}_b\equiv \ket{\alpha}\ket{\alpha}\equiv \ket{\alpha,\alpha} \equiv \ket{\alpha}\otimes \ket{\alpha} $. The application of a creation operator over a coherent state \cite{Agarwal1991} produces the state
\begin{equation}
\ket{\alpha,1}=\frac{\hat{a}^{\dagger}\ket{\alpha}}{\sqrt{1+|\alpha|^2}},
\end{equation}
The states in (\ref{eq:familyofstates}) can be generated through a general beam splitter (BS) operator \cite{Campos1989},
\begin{equation}
\hat{B} (\theta)= \begin{bmatrix}
\cos \theta e^{i\phi_t}  & \sen \theta e^{i\phi_\rho} \\
-\sen \theta e^{-i\phi_\rho} & \cos \theta e^{-i\phi_t}
\end{bmatrix},
\end{equation}
with their inputs either $\ket{\alpha_0}\otimes\ket{1}$ or $\ket{\alpha_0,1}\otimes \ket{0}$. Here, we restrict to a common BS with no phase on transmitted beams and a $\pi$ phase on the reflected ones ($\phi_t=0, \phi_\rho=\pi$), and $\theta=\pi/4$, which represents a 50/50 BS and assume no other phase differences on output beams.

For the negative superposition, the output of the BS is determined by $\ket{\psi}_{\text{dp}}= \hat{B}(\theta=\pi/4) \ket{1}\ket{\alpha}$ before loss and measurement. Equation (\ref{eq:familyofstates}) is reduced to
 \begin{equation}
 	\ket{\psi}_{\text{dp}} =\frac{1}{\sqrt{2}}( \hat{a}^\dagger -  \hat{b}^\dagger )
 	\ket{\alpha }\ket{\alpha},
 	\label{eq:sekgral}
 \end{equation}
 where $\alpha = \alpha_0 /\sqrt{2}$, the state is normalized and dp stands for  displaced photons channel(DPC).

\begin{figure}
\floatbox[{\capbeside\thisfloatsetup{capbesideposition={left,top},capbesidewidth=4cm}}]{figure}[\FBwidth]
{\caption{Generation process using a BS  (a) DPC (\ref{eq:PA1}) and (b) PAC (\ref{eq:sekgral}). The correlations are characterize by means of Homodyne detections.  $\Phi_{\lambda_{A(B)}}$ represents an homodyne measurement with a phase $\lambda_{A(B)}$.}\label{fig:Esquemasimplificado}}
{\includegraphics[width=7cm]{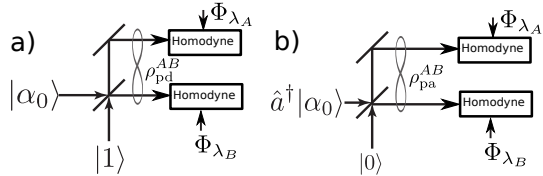}}
\end{figure}

 In the positive superposition, the output of the BS will be $\ket{\psi}_{\text{pa}}= \hat{B}(\theta=\pi/4)  \ket{\alpha,1}\otimes \ket{0}$. The state (\ref{eq:familyofstates}) then reduces to
 \begin{equation}
 	\ket{\psi}_{\text{pa}}=\frac{\mathcal{N}}{\sqrt{2}}( \hat{a}^\dagger + \hat{b}^\dagger)
 	\ket{\alpha }\ket{\alpha},
 	\label{eq:PA1}
 \end{equation}
 where $\mathcal{N}=1/\sqrt{ {}_{\text{pa}}   \bra{\psi}  \psi \rangle_{\text{pa}}   }=1/\sqrt{1+2|\alpha|^2} $ is the normalization constant, $\alpha=\alpha_0/\sqrt{2}$. The subscript pa stands for photon added and designates the input state $\ket{\alpha,1}$ to the BS in Fig. \ref{fig:Esquemasimplificado}(b) that gives origin to (\ref{eq:PA1}). We  call this  a photon added channel (PAC).
Equivalently (\ref{eq:familyofstates}) can be rewritten  in the orthogonal basis,
\begin{equation}
\begin{split}
	\ket{\psi}&=(\mathcal{N}/\sqrt{2})(\hat{D}(\alpha) \ket{1}_a\ket{\alpha}_b \pm \ket{\alpha}_a\hat{D}(\alpha) \ket{1}_b\\
	&\quad+(\alpha^* \pm \alpha^*)\ket{\alpha}_a\ket{\alpha}_b).
	\end{split}
	\label{eq:repalt}
\end{equation}

Specifically, for the dp (\ref{eq:sekgral}) state, it is reduced to
 \begin{equation}
 \begin{split}
\ket{\psi}_{\text{dp}} &=(1/\sqrt{2})( \hat{D}(\alpha_0/\sqrt{2}) \ket{1}_a\ket{\alpha_0/\sqrt{2}}_b \\
 &\quad-  \ket{\alpha_0/\sqrt{2}}_a\hat{D}(\alpha_0/\sqrt{2}) \ket{1}_b ),
\end{split}
\label{eq:dp_esp}
 \end{equation}
 and becomes the state proposed in \cite{Sekatski2012a}. On the other hand, for the state pa (\ref{eq:PA1}), the generated state is
 \begin{equation}
 	\begin{split}
 	\ket{\psi}_{\text{pa}} &= \mathcal{N}( \hat{D}(\alpha_0/\sqrt{2}) \ket{1}_a\ket{\alpha_0/\sqrt{2}}_b\\
 	& \quad +  \ket{\alpha_0/\sqrt{2}}_a\hat{D}(\alpha_0/\sqrt{2}) \ket{1}_b \\
 	& \quad + \sqrt{2}\alpha_0 \ket{\alpha_0/\sqrt{2}}_a \ket{\alpha_0/\sqrt{2}}_b),
 	\end{split}
 	\label{eq:pa_esp}
 \end{equation}
it is easy to see that the term $\sqrt{2}\alpha_0 \ket{\alpha_0/\sqrt{2}}_a \ket{\alpha_0/\sqrt{2}}_b$ plays the role of entanglement loss compared to the single photons entangled state that occurs at $\alpha_0=0$.

Bipartite states $\ket{\psi}_{\text{pd}}$ and $\ket{\psi}_{\text{pa}}$, form a shared resource channel between Alice and Bob, identified by its pure density matrix,
\begin{equation}
	\rho^{AB}_{\text{dp(pa)}}=\ket{\psi}_{\text{dp(pa)  (pa)dp}}\bra{\psi}.
	\label{eq:densitymatrixGral}
\end{equation}

The generation schemes for the pure states to be analyzed are outlined in Fig. \ref{fig:Esquemasimplificado}. In \ref{fig:Esquemasimplificado}(a) a coherent state $\ket{\alpha_0}$ and a fock state $\ket{1}$ are mixed in a 50/50 BS which creates $\rho_{\text{pd}}^{AB}$, and in \ref{fig:Esquemasimplificado}(b) a SPACS $\hat{a}^\dagger \ket{\alpha}$ is mixed with the vacuum $\ket{0}$ and creates $\rho_{\text{pa}}^{AB}$. Both figures include a homodyne detection process as a first approach to analyze correlations of the bipartite channel created by the BS. This proposed homodyne measurement is intended for evaluating only the pure channel correlation  without any source of loss by means of quadrature distribution.
  These measurements can be considered as a signature of non-classicality measured by the POVM (\ref{eq:povm_hom}). The joint probability of the quadratures $X_{\lambda_A}$ and $X_{\lambda_B}$, when a homodyne detection is done with phases $\lambda_A$ and $\lambda_B$ respectively, and is calculated using (\ref{eq:jointprobquad}) with the measurement scheme shown in Fig. \ref{fig:Esquemasimplificado}. MID will also be used to characterize the correlation present in the pure entangled channel.

\begin{figure}
\floatbox[{\capbeside\thisfloatsetup{capbesideposition={left,top},capbesidewidth=4cm}}]{figure}[\FBwidth]
{\caption{Schematic characterization of the entangled channel by simultaneous detections. The input state is either the DFS forming the DPC shown in (a), or SPACS forming the PAC shown in (b). Loss mechanisms are included as a fluctuating phase $\phi$ and scattering as a BS with transmittance $\eta$. Bob side is measured after a displacement $\hat{D}(-\alpha)$ by $M_i^B$ and Alice located at long distance measures the state received by $M_i^A.$}\label{fig:esquema}}
{\includegraphics[width=7cm]{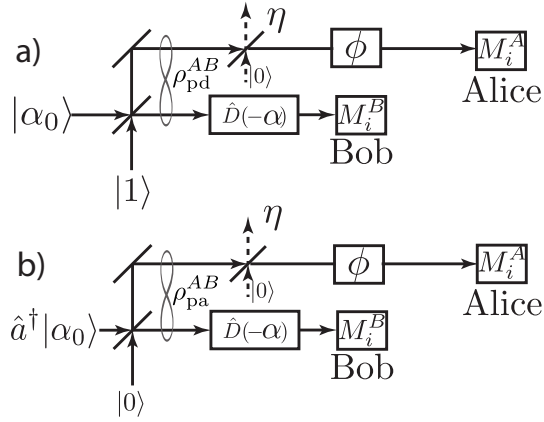}}
\end{figure}

In Fig. \ref{fig:esquema}, we present a simple strategy to use the bipartite channel to exploit its correlations between two distant parties, where loss mechanisms have been included for both states $\rho_{\text{pd}}^{AB}$ and  $\rho_{\text{pa}}^{AB}$ shown in  \ref{fig:esquema}(a) and \ref{fig:esquema} (b), respectively. The loss sources are modeled in the channels as follows: for the scattering, with a BS coupled with the vacuum with transmittance $\eta$, and for the phase fluctuation a phase difference $\phi$ between upper and lower arms as is shown in Fig. \ref{fig:esquema} (a)-(b). It is considered that Bob creates the resource state and keeps his side to perform some measurements and sends the other part to Alice remotely located, and Alice wants to gain information on the measurement performed by Bob. A displacement $\hat{D}(-\alpha)$ prior to Bob measure has been included, whose purpose is that Bob obtains his state in the standard qubit basis, instead of a continuous state. Displacement is the unitary operator $\hat{D}(\alpha)= e^{\alpha \hat{a}^\dagger}-e^{\alpha^* \hat{a}}$, such that if it acts on the vacuum creates a coherent state $\hat{D}(\alpha) \ket{0}=\ket{\alpha}$.

By considering $\hat{a}$ the input mode and $\hat{c}$ the vacuum input, transmittance is parameterized as $\eta=\cos^2\theta'$, where $\theta'=\arccos (\sqrt{\eta})$ with $\phi_t=0$ and $\phi_\rho=\pi$. The scattering transform the pure state (\ref{eq:densitymatrixGral}) following $\hat{a}^{'}=\sqrt{\eta}\hat{a}-\sqrt{1-\eta}\hat{c}$ and $\hat{c}^{'}=\sqrt{\eta}\hat{a}+\sqrt{1-\eta}\hat{c}$, where $a'(c')$ represent the output modes of the BS \cite{leonhardt2003,Leonhardt2010,Sekatski2012a}. This transforms the state (\ref{eq:familyofstates}) into a tripartite state

\begin{equation}
\begin{aligned}
&\ket{\psi(\eta)}_{\text{pd(pa)}}=\hat{B}(\theta ') \ket{\psi}_{\text{pd(pa)}}\\
&\quad= \mathcal{N}^2 ( \frac{\sqrt{\eta}}{\sqrt{2}}\hat{D}_{a'}(\alpha \sqrt{2})\hat{D}_{c'}(\sqrt{1-\eta}\alpha)\ket{1}_{a'}\ket{0}_{c'}\ket{\alpha}_{b}\\
&\quad+\frac{\sqrt{1-\eta}}{\sqrt{2}}\hat{D}_{a'}(\alpha \sqrt{\eta})\hat{D}_{c'}(\sqrt{1-\eta}\alpha)\ket{0}_{a'}\ket{1}_{c'}\ket{\alpha}_{b}\\
&\quad \mp\frac{1}{\sqrt{2}} \hat{D}_{a'}(\alpha \sqrt{\eta})\hat{D}_{c'}(\sqrt{1-\eta}\alpha)\hat{D}(\alpha)_{b}\ket{0}_{a'}\ket{0}_{c'}\ket{1}_{b}\\
&\quad +\frac{\alpha^*}{\sqrt{2}}(1\mp 1) \hat{D}_{a'}(\alpha \sqrt{\eta})\hat{D}_{c'}(\sqrt{1-\eta}\alpha) \ket{0}_{a'}\ket{0}_{c'}\ket{\alpha}_{b}).
\end{aligned}
\end{equation}
The density matrix of this  system, including loss,  is $\rho_{\text{pd(pa)}}^{AB,loss}(\eta)= \ket{\psi (\eta)}_{\text{pd(pa) (pa)pd}}\bra{\psi(\eta)}$. The final effective state subjected to the scattering  results after a sum over the loss modes ($\hat{c}'$ modes) produces the mixed state,
\begin{subequations}
\begin{equation}
\rho^{AB}_{\text{pd(pa)}} (\eta) =\text{Tr}_\text{loss}[ \hat{B}(\theta') \rho^{AB} \hat{B}^\dagger(\theta')],
\label{eq:mixed1gral}
\end{equation}
which can be rewritten as the mixture
\begin{equation}
\rho_{\text{pd(pa)}}^{AB}(\eta)=p_1 \rho_1(\eta) + p_2 \rho_2(\eta).
\label{eq:mixed2gral}
\end{equation}
\label{eq:mixedgral}
\end{subequations}
This emerges because $\hat{c}'$ modes can be either $\ket{\sqrt{1-\eta}\alpha}\bra{\sqrt{1-\eta}\alpha}$ or $\hat{D}(\sqrt{1-\eta}\alpha)\ket{1}\bra{1}\hat{D}^\dagger(\sqrt{1-\eta}\alpha)$.  where the states  are
\begin{subequations}
	\begin{equation}
		\rho_1(\eta)= \hat{D}(\alpha \sqrt{\eta})\hat{D}(\alpha) \ket{00}\bra{00}\hat{D}^\dagger(\alpha)\hat{D}^\dagger(\alpha \sqrt{\eta}),
	\end{equation}
	\begin{equation}
	\begin{aligned}
		\rho_2(\eta)&=(1/(|\alpha|^2(1\mp 1)^2 + 1 +\eta))\hat{D}(\alpha\sqrt{\eta})\hat{D}(\alpha)\\
		&\quad \times [(\alpha^*  \mp \alpha^*)\ket{00} \mp \ket{01} \\
		&\quad+ \sqrt{\eta}\ket{10}   ] \times [\bra{00}(\alpha \mp \alpha)\\
		&\quad \mp \bra{01}+\bra{10} \sqrt{\eta}  ]\hat{D}^\dagger(\alpha)\hat{D}^\dagger(\alpha\sqrt{\eta}),
	\end{aligned}
	\end{equation}
	\label{eq:edosmezcla}	
\end{subequations}

the sign $``-",``+"$ corresponds to pd and pa respectively, where the probabilities are $p_1= (\mathcal{N}^2/2)(1-\eta)$ and $p_2=(\mathcal{N}^2/2)(|\alpha|^2(1\mp 1)^2 +1+ \eta) $. The probabilities have been written for both pd(pa) cases, just note that $\mathcal{N}=1$ for the pd case.

Considering a free space channel, unavoidable refractive index fluctuations will induce arbitrary phase changes. These fluctuations will be modeled as an ensemble with a Gaussian probability distribution over the phase. The arm sent to Alice is assumed to acquire the phase difference $\phi$ with respect to Bob's arm and is included by the action of the relative phase shifting operator $\hat{U}(\phi)=e^{i\phi \hat{a}^\dagger \hat{a} }$ acting over $\hat{a}$ mode as $\rho=\int d\phi p(\phi) \hat{U}(\phi) \rho^{AB} \hat{U}^\dagger (\phi)$. However, the phase fluctuations will be added to the channel already subjected to scattering, leading to the form,
\begin{equation}
\rho_{\text{pd(pa)}} (\eta,\phi) = \int d\phi p(\phi) \hat{U}(\phi)  \rho_{\text{pd(pa)}}^{AB} (\eta)      \hat{U}^\dagger (\phi),
\label{eq:phase1}
\end{equation}
 where $\rho_{\text{pd(pa)}}^{AB}(\eta)$ was defined through (\ref{eq:mixedgral})-(\ref{eq:edosmezcla}), $p(\phi)$ is a normalized Gaussian distribution with variance of phase $\sigma$ centered at zero. The combined effect of the phase and scattering noise is given by
 \begin{equation}
 \hat{U}(\phi)  \rho_{\text{pd(pa)}}^{AB} (\eta)      \hat{U}^\dagger (\phi)= p_1 \rho_1(\eta,\phi)+p_2 \rho_2(\eta,\phi),
 \end{equation}
 where each density matrix in the mixture is as follows,
\begin{subequations}
\begin{equation}
\rho_1(\eta,\phi)=\hat{D}(\alpha e^{i\phi} \sqrt{\eta})\hat{D}(\alpha) \ket{00}\bra{00}\hat{D}^\dagger(\alpha)\hat{D}^\dagger(\alpha e^{i\phi}\sqrt{\eta}),
\end{equation}
\begin{equation}
\begin{aligned}
\rho_2(\eta,\phi)&=(1/2)\hat{D}(\alpha e^{i\phi} \sqrt{\eta})\hat{D}(\alpha) [(\alpha^* \mp \alpha^*)\ket{00}\\
& \mp \ket{01} + \sqrt{\eta}e^{i\phi}\ket{10}   ] \times [\bra{00}(\alpha \mp \alpha)\\
& \mp\bra{01} +\bra{10}\sqrt{\eta} e^{-i\phi}     ]\hat{D}^\dagger(\alpha)\hat{D}^\dagger(\alpha e^{i\phi}\sqrt{\eta}),
\end{aligned}
\end{equation}	
\label{eq:edosmezcla_etaphi}
\end{subequations}
with the same  $p_1$ and $p_2$ as in (\ref{eq:mixedgral}).

\section{Results}

\subsection{Homodyne Conditional Measurements.} Joint Quadrature Probability (JQP) distribution (\ref{eq:jointprobquad}) is not just a analytical calculation, but an easily accessible experimental measure as well. JPQ, Eq. (\ref{eq:jointprobquad}) for DPC and PAC are obtained as,

	\begin{equation}
	\begin{aligned}
	P&^{AB}_{\text{pd(pa)}}(X_{\lambda_A},X_{\lambda_B})=\frac{(a_1\pm a_2)}{4\pi (1+\frac{1}{2}(1\mp 1)\alpha_0^2)}\\
	&\times \exp \left( -(X_{\lambda_A}- \alpha_0\cos \lambda_A)^2 - (X_{\lambda_B}-\alpha_0\cos \lambda_B)^2 \right),
	\end{aligned}
	\label{eq:JQPPD}
	\end{equation}
	where
	\begin{equation*}
	\begin{aligned}
	a_1&=4| e^{i \lambda _A}(X_{\lambda _A}-\frac{1}{2}e^{-i \lambda _A}\alpha_0^2) |^2\\
	&\quad +4| e^{i \lambda _B}(X_{\lambda _B}-\frac{1}{2}e^{-i \lambda _B}\alpha_0^2) |^2,\\
	a_2&=2 \Re \Big(e^{i(\lambda_B-2\lambda_A)}\left(-\alpha_0+2e^{i\lambda_A}X_{\lambda_A}\right)\Big)\\
	&\quad \times \left( -2X_{\lambda_B}+\alpha_0 e^{i\lambda_B}   \right).
	\end{aligned}
	\end{equation*}
From (\ref{eq:JQPPD}) it is clear that JQP for both cases is a Gaussian function in the two variables $X_{\lambda_{A}}$ and $X_{\lambda_{B}}$ times a polynomial function of the quadratures. The Gaussian functions are not centered at the origin; instead, their center positions depend on the phase values $\lambda_A(\lambda_B)$ and the value of $\alpha_0$ of the coherent state. Also, the quadrature correlation is contained in the term $a_2$.

\begin{figure}
\floatbox[{\capbeside\thisfloatsetup{capbesideposition={left,top},capbesidewidth=4cm}}]{figure}[\FBwidth]
{\caption{JQP for the pure DPC with homodyne detectors for different values of $n_0$. The parameters for each plot are (a) $\lambda_A=0,\lambda_B=0,n_0=0$, (b) $\lambda_A=0,\lambda_B=0,n_0=81$, (c) $\lambda_A=\pi/2,\lambda_B=0,n_0=0$, (d) $\lambda_A=\pi/2,\lambda_B=0,n_0=81$.}\label{fig:exp7}}
{\includegraphics[width=7cm]{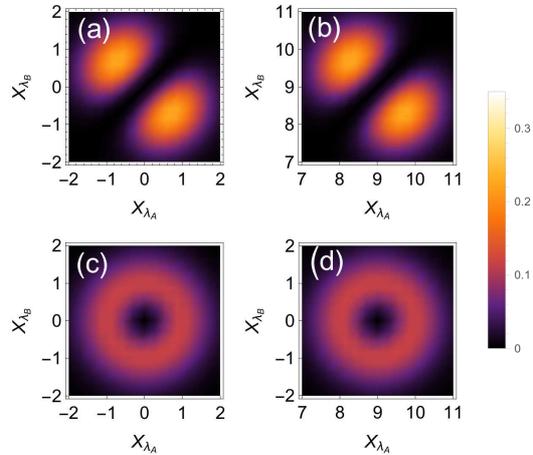}}
\end{figure}

JQP for the pure state $\rho^{AB}_{\text{pd}}$, Eq. (\ref{eq:JQPPD}), is shown  in Fig. \ref{fig:exp7} for two configurations of the local oscillator phases and two combinations of the average photon numbers. In Fig. \ref{fig:exp7}(a) and (b) the detectors are set to phase $\lambda_A = 0$ and $\lambda_B=0$ with photon number $n_0=0$ and $n_0=81$, respectively. The Fig. \ref{fig:exp7}(c) and (d) is for the cases  $\lambda_A = \pi/2$ and $\lambda_B=0$ with $n_0=0$ and $n_0=81$, as well. Its easily seen that quadrature correlations remain unchanged for different $n_0$. One can note that quadratures exhibit a highly correlated behavior, since for (a) and (b) the values for $X_{\lambda_A}$ and $X_{\lambda_B}$ are correlated in the illuminated regions, while in (c) and (d) for a given value of $X_{\lambda_A}$ a set of values of $X_{\lambda_B}$ is more likely to be found in the ring symmetry shown. The JQP behavior for $\rho^{AB}_{\text{pd}}$ is independent of the average photon number.

\begin{figure}
\floatbox[{\capbeside\thisfloatsetup{capbesideposition={left,top},capbesidewidth=4cm}}]{figure}[\FBwidth]
{\caption{JQP for the pure PAC with homodyne detectors for different values of $n_0$. The parameters for each plot are: (a) $\lambda_A=0,\lambda_B=0,n_0=0$, (b) $\lambda_A=0,\lambda_B=0,n_0=1$, (c) $\lambda_A=\pi/2,\lambda_B=0,n_0=0$, (d) $\lambda_A=\pi/2,\lambda_B=0,n_0=1$.}\label{fig:exp6}}
{\includegraphics[width=7cm]{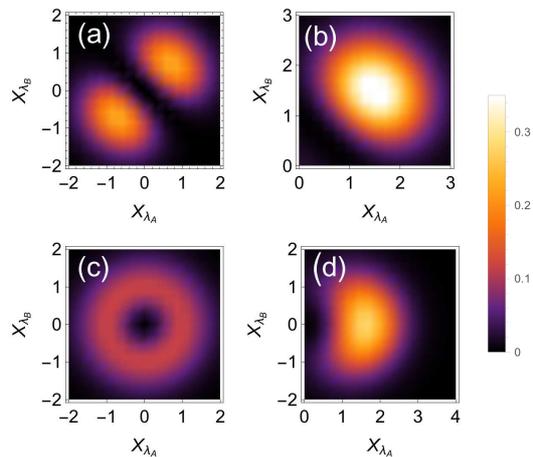}}
\end{figure}

JQP  distribution is also calculated for the pa pure state $\rho^{AB}_{\text{pa}}$ (\ref{eq:JQPPD}) and is shown in Fig. \ref{fig:exp6} with same configuration of the phase detectors but with the average photon numbers as $n_0=0$ in (a) and (c), and $n_0=1$, in (b) and (d). The behaviour shown Fig. \ref{fig:exp6}(a) and (c) exhibit a similar correlated behavior as for the $\rho^{AB}_{\text{pa}}$ simply because if $n_0=0$ the state is virtually the same up to a sign in the superposition, what becomes evident in the $\pi/2$ rotated correlation shown in \ref{fig:exp6}(a)  when compared to Fig. \ref{fig:exp7}(a). It is easily seen that as the average photon number increases, quadrature correlations drop rapidly as shown in \ref{fig:exp7}(b) and (d) even for the low value $n_0=1$, where the density plots show low correlation between the values of $X_{\lambda_A}$ and $X_{\lambda_B}$. No further average photon numbers were considered for this case because the JQP gets rounded quickly, which is a signature of separable and uncorrelated states.

By analyzing the JQP  we can then have a simple approach to know and test if a bipartite state has some degree of correlation. Therefore, the finding of some  correlations for a pair of homodyne detection phases $\lambda_{A,B}$, is a starting point to study of the quantum correlations of the state.

\subsection{\label{MIDpure}MID for the pure channel.}

We now  analyze the pure channel properties with the evaluation of  the MID (\ref{eq:MIDdef}) by using (\ref{eq:MID6})-(\ref{eq:MID5}) in the entropy calculation as a first approach to understand how correlations behave for different phases of the quadrature detections. Even when the MID calculation is carried out on the pure channel, it has interesting implications since it can easily show how to obtain maximum correlations for the selected measurement strategy. It shows the relationship and values between  the homodyne phases $\lambda_A$ and $\lambda_B$ that accomplishes this. The entropies can be obtained from a subsystem as $S(\rho^{A}_{\text{pd(pa)}})=-\rho^A_{\text{pd(pa)}} \log \rho^A_{\text{pd(pa)}}$, and gives,
\begin{subequations}
	\begin{equation}
		S(\rho^A_{\text{pd}})=1,
	\end{equation}
	\begin{equation}
		S(\rho^A_{\text{pa}})=\frac{ \left(\ln \left(2 \left(\alpha _0^2+1\right)\right)-\frac{\alpha _0 \sqrt{\alpha _0^2+2} \tanh ^{-1}\left(\frac{\alpha _0 \sqrt{\alpha _0^2+2}}{\alpha _0^2+1}\right)}{\alpha _0^2+1}\right)}{\ln (2)},
	\end{equation}
	\label{eq:entVNrhoA_PDPA}
\end{subequations}
that will be used for the  MID and AMID calculations.

\begin{figure}
	\centering
	\subfloat[]{\includegraphics[width=0.3\textwidth]{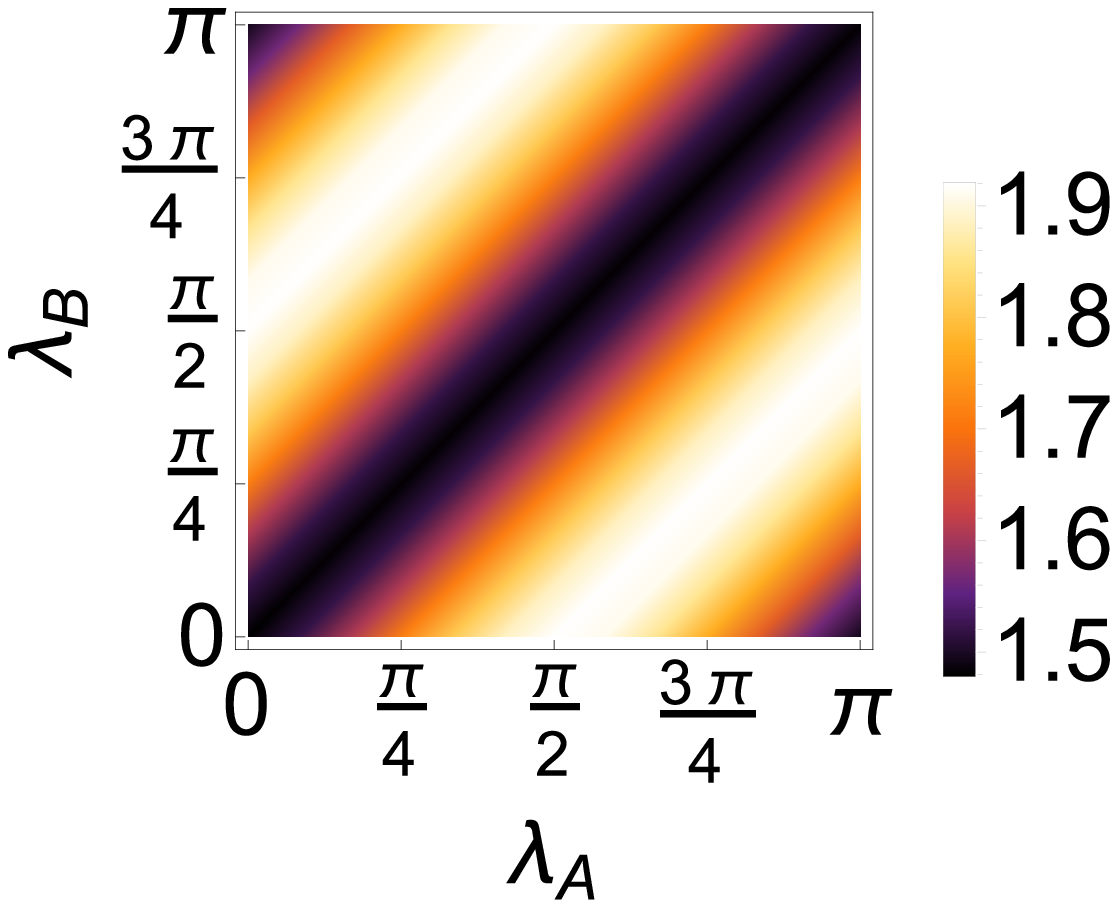}}
	\subfloat[]{\includegraphics[width=0.3\textwidth]{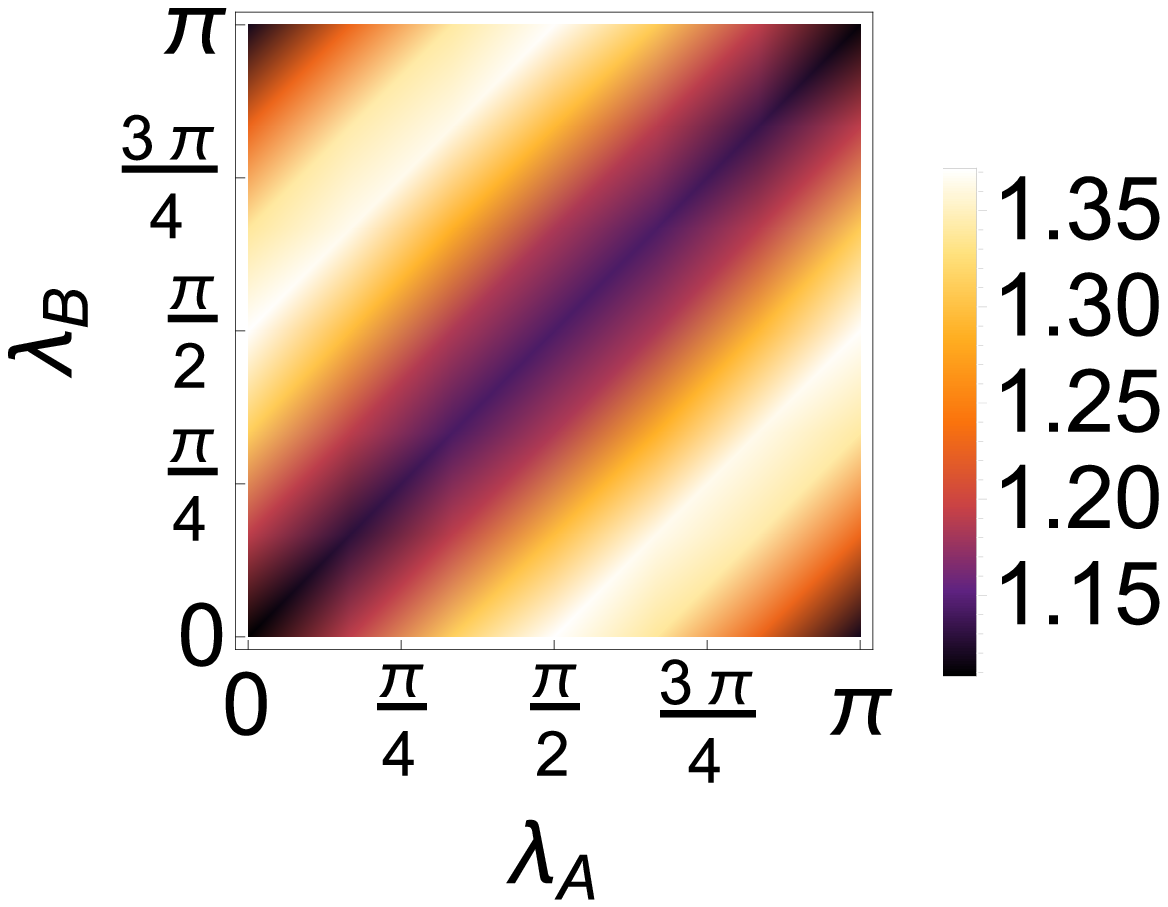}}\\
	\subfloat[]{\includegraphics[width=0.3\textwidth]{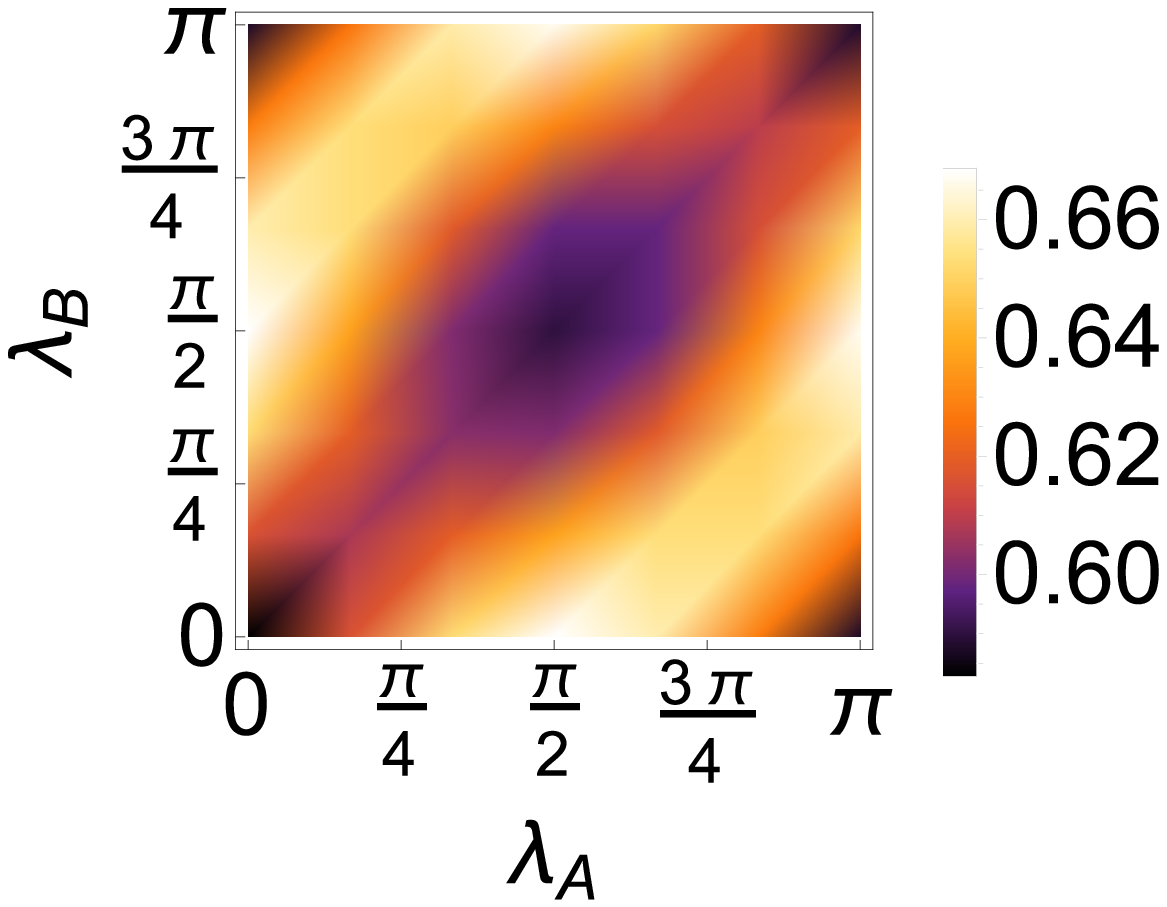}}
	\subfloat[]{\includegraphics[width=0.3\textwidth]{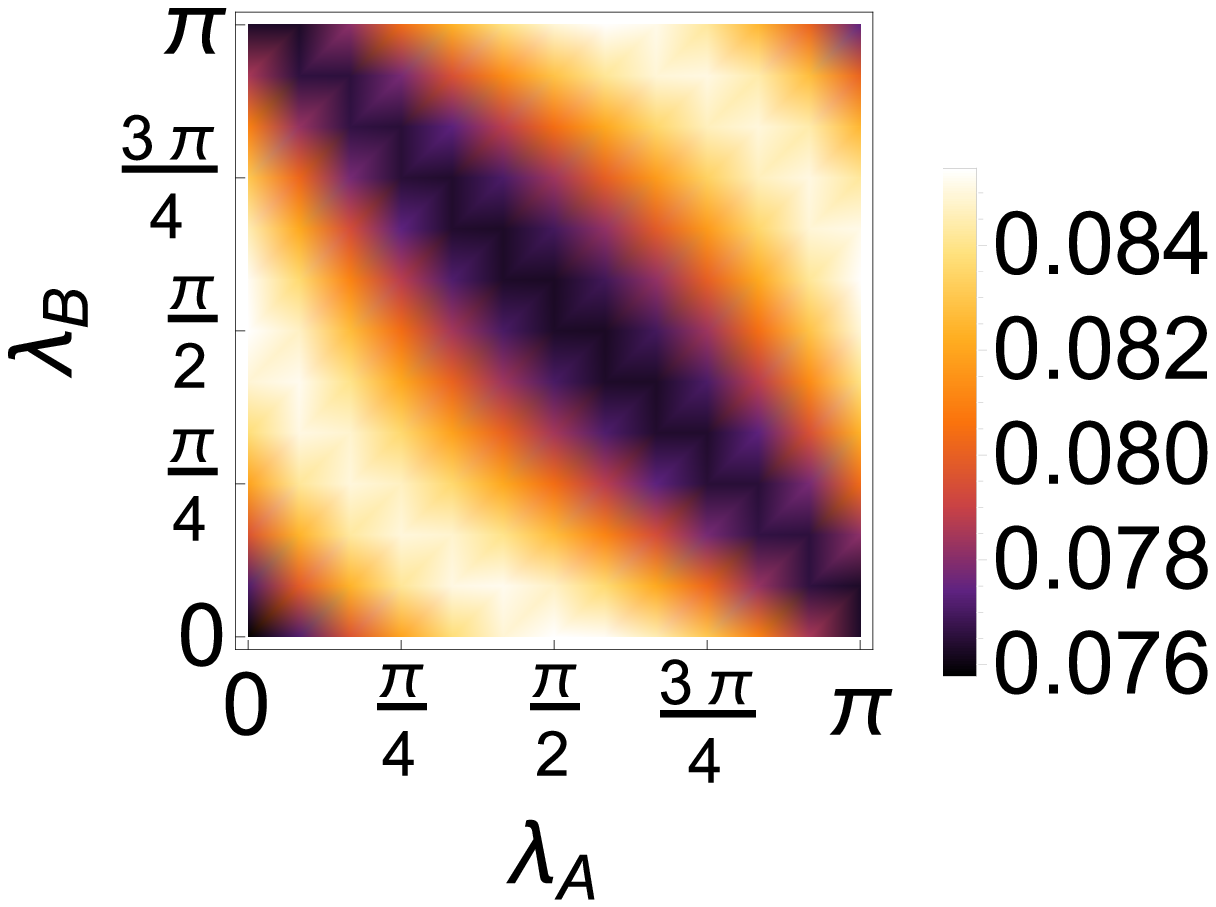}}
	\caption{MID dependent of the phase detectors $\lambda_A$ and $\lambda_B$ vs. $n_0$ (a) MID for DPC independent of $n_0$. MID for PAC with parameters: (b) $n_0=0.25$, (b) $n_0=1$ (c) $n_0=6.25$.}
	\label{fig:MIDmap}
\end{figure}

\begin{figure}
\floatbox[{\capbeside\thisfloatsetup{capbesideposition={left,top},capbesidewidth=5cm}}]{figure}[\FBwidth]
{\caption{Comparison of entropy of entanglement and AMID for pure PAC and DPC.}\label{fig:entvsamid}}
{\includegraphics[width=6cm]{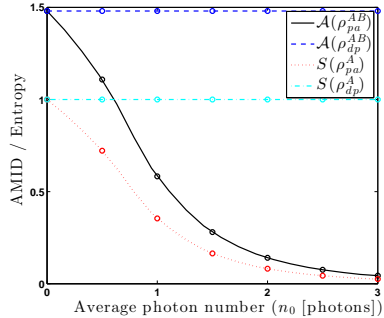}}
\end{figure}

Results for MID calculation for both channels are shown in Fig. \ref{fig:MIDmap}. In Fig. \ref{fig:MIDmap}(a) we present  the MID for the state $\rho_{\text{dp}}$ valid for all $n_0$, whereas in Fig. \ref{fig:MIDmap}(b)-(d) we show the MID for the state $\rho_{\text{pa}}$ for average photon number $n_0=\{0.25,1,6.25\}$ respectively. Intensity maps show  higher correlations for the brighter intensities and vice-versa, i.e., that darker regions imply that measurement induced classical correlations are lower. From Fig.\ref{fig:MIDmap}(a) it is evident that for $\rho^{AB}_{\text{pd}}$ case, the maximal correlations occur with a dephasing of $\pi/2$ (the lightest value in the figure). For the case $\rho^{AB}_{\text{pa}}$, in Figs. \ref{fig:MIDmap}(b) to (d) show that as $n_0$ increases, classical correlations increase as well and MID decreases, also a fading out behavior is observed along with a rotation counter clock wise of the density plots. Fig.\ref{fig:MIDmap}(a) shows that minimum of MID for $\rho^{AB}_{\text{pd}}$ occurs for all $\lambda_A=\lambda_B$. Meanwhile for $\rho^{AB}_{\text{pa}}$, minimum of MID  depends highly on $n_0$. However $S(\rho^A)+S(\rho^B)$ is maximal when $\lambda_A=\lambda_B=0$, and at the same time $S(\rho^{AB})$ is minimal for the same $\lambda_{A(B)}$ values. This is confirmed with the numerical calculation shown in Figs. \ref{fig:MIDmap}(b)-(d). These conditions lead to simplified expressions shown in appendix \ref{app1},  that allow  the full calculation of AMID shown in Fig. \ref{fig:entvsamid}.

AMID (\ref{eq:AMIDdef}) is also evaluated and its behaviour is shown in Fig. \ref{fig:entvsamid}. The minimum, is calculated numerically for each point shown. The AMID $\mathcal{A}(\rho^{AB}_{\text{pd(pa)}})$ and entropy of one subsystem $\rho^A$ $S(\rho^A_{\text{pd(pa)}})$ is shown vs. the average photon number $n_0$. The entropies from $\mathcal{I}(\rho^{AB}_{\hat{\Pi}})$ are $\{\lambda_A,\lambda_B\}$ dependent and obtained from Eq. (\ref{eq:I_proy}). Entropies are obtained by using  the reduced density matrix $\rho^A$ (equivalently $\rho^B$ can be selected as it is a pure state) and AMID by (\ref{eq:AMIDdef}) with the optimal angles discussed above with the help of the  expressions in appendix \ref{app1}. The dot marks are calculated values and lines are an interpolation. Entropy and AMID are constant values for the state $\rho^{AB}_{\text{pd}}$ what is in agreement with the maximally entangled nature of the system. On the contrary, entropy and AMID for  $\rho^{AB}_{\text{pa}}$ are highly dependent on the average photon number  $n_0$, and decay to zero for low values of $n_0$. From Fig. \ref{fig:entvsamid} one can see that the obtained behavior by the quadrature AMID proposed here is qualitatively the same as the one obtained from the entropy, but in a different scale, what reinforces its practical utilization for characterizing quantum correlations in  continuous bipartite states.

In an attempt to find an AMID analytical expression for the pure case, the behavior of $\mathcal{A}(\rho^{AB})$ is analyzed, which will be minimum when $S(\rho^A)+S(\rho^B)$ is maximum and $S(\rho^{AB})$ is minimum. Maxima and minima of the phase dependent entropies are obtained by inspection of the MID numerical behavior shown in Fig. \ref{fig:MIDmap} where AMID can be identified with the black values, as they represent the minimum of the MID. This graphical analysis for the minimization is followed instead of gradient analysis because of the lengthy nature of the expressions, and that the simple selection of $\lambda_A=\lambda_B=0$ works for both studied cases giving the minimum of MID. The AMID for the $\rho_{\text{pa}}$ case as a function of $n_0$ and can be fitted to the following function
\begin{equation}
\mathcal{A}(\rho_{pa}^{AV})=1/(a+\exp (b (n_0-c)),
\label{eq:FitAMID}
\end{equation}
where $a=0.51262 \pm 0.0511021$, $b= 1.95072 \pm 0.211388$ and $c= 0.94674 \pm 0.0491102$. The same qualitative behavior of entropy and AMID is evident from the fact that if (\ref{eq:FitAMID}) is normalized, overlaps to $S(\rho_{pa}^{A})$,  what also establishes an appropriate fit for its entropy with (\ref{eq:FitAMID}) rescaled. See appendix \ref{app1} for further details, where $\lambda_{A(B)}$ dependent entropies are shown.

\subsection{Quantum Discord: quantitative correlations}
\subsubsection{Loss by scattering}
In this section, the effect on quantum and classical correlations is studied quantitatively for both cases, where  we now include a scattering source of loss as in (\ref{eq:mixed1gral}) by means of determination of the Quantum Discord from B to A. To accomplish this, we follow, equation (\ref{eq:phase1}) that describes the bipartite noise dependent channel. QD from B to A requires to account for the effect of measuring B with a complete set of POVM elements. Given the hybrid nature of the channel, the projective measure is chosen by a POVM with operators
\begin{equation}
\hat{\Pi}_i^B \equiv\hat{D}(\alpha)\ket{i_M}_{BB}\bra{i_M}\hat{D}^\dagger(-\alpha),
\label{eq:POVM_disp}
\end{equation}
where $\ket{i_M}_{BB}\bra{i_M}$ are the standard 1-qubit POVM elements, with $\ket{0_M}=\cos (\theta_M/2)\ket{0}+e^{i\phi_M}\sin{\theta_M/2}$, and $\ket{1_M}=\sin(\theta_M/2)\ket{0}-e^{i\phi} \cos(\theta_M/2)\ket{0} $. This process can be achieved by first displacing $\hat{D}(-\alpha)$ and then measuring with the standard qubit POVM elements as shown in the figure \ref{fig:esquema}. The displacement can be easily accomplished in a laboratory as shown by M.G.A. Paris \cite{Paris1996}. Following this, its guaranteed that Bob's subsystem is fully measured and the  quantum discord calculation is feasible in this continuous model.

\textbf{Superposition of DFS: DPC}. First, the QD for the  state $\rho^{AB}_{\text{dp}}$ is analyzed, following (\ref{eq:QD}). To do this, the state is projected by (\ref{eq:POVM_disp}) which creates entropy expressions that depend on $\alpha_0,\eta,\theta_M, \phi_M$, where $\alpha_0$ is the amplitude of the CS used to create the superpositions under study and $\eta$ is the scattering effect of the channel as depicted in Fig.\ref{fig:esquema}, $\theta_M$ and $\phi_M$ are the projection angles of the measurements (\ref{eq:POVM_disp}). To obtain a closed analytical expression for the QD dependent on $(\alpha_0,\eta,\theta_M, \phi_M)$ is a complex task. As $S(\rho^B)-S(\rho^{AB})$ is only $\eta$ dependent, we focus on the term $\underset{\{\Pi_i\}}{\text{min}} [S(\rho^A_{\Pi^B})(A|B)]$. To try to overcome this difficulty we calculate $S(\rho^A_{\Pi^B})(A|B)$ for different values of $\eta$ aiming to identify the behavior of the measuring angles and its optimization. We observe that all the classical correlations obtained are non $\phi_M$ dependent, and then $\partial_{\theta_M} S(\rho^A_{\Pi^B})(A|B)=0$ occurs at the angles $\theta_M^*=\{0,\pi/2\}$. The minimization problem gets reduced to only two discrete angles. The global minimum is found for $\theta_M=\pi/2$. As long as the optimal measurements angles are determined, an expression independent of $\alpha_0$ for $S(\rho^A_{\Pi^B})(A|B) (\eta)$ can be easily obtained.

With previous considerations, we obtain the reduced entropy $S(\rho^B)=\sum_i \lambda_i^B \log \lambda_i^B$ and $S(\rho^{AB})= \sum_i \lambda_i^{AB} \log \lambda_i^{AB}$ with $\lambda_{i}^B=\{\frac{1}{2},\frac{1}{2}\}$ and the full entropy $\lambda_{i}^{AB}=\{0,0,\frac{1-\eta}{2},\frac{1+\eta}{2}\}$. The optimized conditional entropy is reduced to $S(\rho^A_{\Pi^B})(A|B) = p_1 S(\rho_{\Pi_1^B}^{A})+ p_2 S(\rho_{\Pi_2^B}^{A})$, where $S(\rho_{\Pi_1}^{A})=\sum_i\lambda_i^{A_{\Pi_1}} \log \lambda_i^{A_{\Pi_1}} $ and $S(\rho_{\Pi_2}^{A})=\sum_i\lambda_i^{A_{\Pi_2}} \log \lambda_i^{A_{\Pi_2}} $ , with the correspoonding eigenvalues $\lambda_i^{A_{\Pi_1}}=\lambda_i^{A_{\Pi_2}}=\frac{1}{2}(1\pm \sqrt{1+\eta(\eta-1)}) $ and probabilities $p_1=p_2=\frac{1}{2}$. The QD for $\rho^{AB}_{\text{pd}}$ under noise by scattering is summarized as,
\begin{equation}
\begin{split}
D_B(\rho^{AB}_{\text{dp}})&=\frac{\ln \left(\frac{4}{\eta }+4\right)+2 \eta  \tanh ^{-1}(\eta )}{\ln (4)}\\
	&\quad +\frac{-2 \sqrt{(\eta -1) \eta +1} \tanh ^{-1}\left(\sqrt{(\eta -1) \eta +1}\right)}{\ln (4)}.
\end{split}
\label{eq:QDwithETAforDP}
\end{equation}
One can note the limits $D_B(\rho^{AB}_{\text{dp}}) \rightarrow 0$ when $\eta \rightarrow 0$ because zero transmissivity avoids the generation of any bipartite state, and  $D_B(\rho^{AB}_{\text{dp}}) \rightarrow 1$ when $\eta \rightarrow 1$ because the initial state is pure and maximally entangled. Alice, who presumably is located far away from Bob, wants to gain information on the QD of the channel they share, and given the continuos variable and mixed nature of the state received, a local homodyne detection is a viable option. The variance of quadrature on her side $\Delta X_{\lambda_A}^2= \braket{ \hat{X}_{\lambda_A}^2}-\braket{\hat{X}_{\lambda_A}}^2$ is an easily accessible measure, with the average value of the quadrature given by $\braket{\hat{X}_{\lambda_A}}=\text{Tr} ( \hat{X}_{\lambda_A} \rho^A) $. Without loss of generality, we assume  $\alpha_0$ real, which produces $\Delta X_{\lambda_A}^2=\frac{1}{2}(1+\eta)$. The variance can be inverted to obtain a QD expression related to experimental data of quadrature from (\ref{eq:QDwithETAforDP}),
\begin{equation}
\begin{split}
D_B(\rho^{AB}_{\text{dp}})&=\frac{1}{\ln 4}\ln \left(\frac{8 \Delta X_{\lambda_A}^2 }{2 \Delta X_{\lambda_A}^2 -1}\right) -\frac{2d_1^{\text{dp}}\tanh ^{-1} d_1^{\text{dp}}}{\ln 4}\\
&\quad+\frac{d_2^{\text{dp}}\tanh ^{-1} (d_2^{\text{dp}}/2)}{\ln 4},
\end{split}
\label{eq:QDdf}
\end{equation}
where $d_1^{\text{dp}}=\sqrt{4 \Delta  X_{\lambda_A}^4-6 \Delta  X_{\lambda_A}^2 +3}$ and $d_2^{\text{dp}}=2-4\Delta X_{\lambda_A}^2$. This is a key result and contribution of this work, as it relates QD of a maximally entangled channel subjected to loss by scattering as a function of the quadrature $\Delta X_{\lambda_A}$, an observable of one subsystem.

The expressions in (\ref{eq:QDwithETAforDP}) and (\ref{eq:QDdf}) are analytical expressions for QD for the DPC channel. It is important to note that, given the independence with respect to the average photon number $n_0$ of this particular channel, it also represents QD for a channel formed by the Bell state $\ket{\psi^{-}}=(1/\sqrt{2}) (\ket{01}-\ket{10}) $ with scattering noise. Therefore (\ref{eq:QDdf}) relates an observable to that degree of quantum correlation, which to our knowledge has not been obtained before.

\textbf{Superposition of SPACs: PAC}. The amount of achievable information by a measure of QD for the initial shared state $\rho^{AB}_{\text{pa}}$ is outlined in a similar way as was done for the  state $\rho^{AB}_{\text{dp}}$. As the matrix of the mixed state contains more terms, and is no longer self normalized (i.e., $\mathcal{N}=1/(1+|\alpha_0|^2)$), it is more difficult than before to obtain a closed expression for the conditional entropy $S(\rho^A_{\Pi^B})(A|B)$, which depends on all parameters $(\alpha_0,\eta,\theta_M,\phi_M)$. Without a closed analytical expression to work with, and in order to solve the minimization problem of $S(\rho^A_{\Pi^B})(A|B)$ with multiple variables, we perform the search of the angles $\theta_M^{*}, \phi_M^*$ numerically for different combinations of $\alpha_0, \eta$. Our numerical analysis  shows that all the minima coincide with  $\theta_M^*=\pi/2, \phi_M^*=\pi/2$. To calculate the discord $D_B(\rho^{AB}_{\text{pa}})$  we proceed as before, with the eigenvalues of $\rho_{\text{pa}}^{AB}, \rho_{\text{pa}}^{B}$ and the projected matrices $\rho_{\Pi_1}^{A}$ and $\rho_{\Pi_2}^{A}$, which are, $\lambda_{i}^{AB}=\{0,0,\frac{\text{$\alpha_0 $}^2-\sqrt{\text{$\alpha_0 $}^4+2 \text{$\alpha_0 $}^2+\eta ^2}+1}{2 \left(\text{$\alpha_0 $}^2+1\right)},\frac{\text{$\alpha_0 $}^2+\sqrt{\text{$\alpha_0 $}^4+2 \text{$\alpha_0$}^2+\eta ^2}+1}{2 \left(\text{$\alpha_0$}^2+1\right)}\}$, and $\lambda_{i}^{B}= \{ \frac{\alpha _0^2-\sqrt{\alpha _0^4+2 \alpha _0^2}+1}{2 \left(\alpha _0^2+1\right)},\frac{\alpha _0^2+\sqrt{\alpha _0^4+2 \alpha _0^2}+1}{2 \left(\alpha _0^2+1\right)} \} $. As in the previous case, the minimal conditional entropy occurs when both possible projections have the same eigenvalues $\lambda_i^{A_{\Pi_1}}=\lambda_i^{A_{\Pi_2}} \{\frac{1}{2}-\frac{\sqrt{\alpha _0^4+2 \alpha _0^2+(\eta -1) \eta +1}}{2 \left(\alpha _0^2+1\right)},\frac{1}{2} \left(\frac{\sqrt{\alpha _0^4+2 \alpha _0^2+(\eta -1) \eta +1}}{\alpha _0^2+1}+1\right)  \} $ and equal probabilities  $p_{1}=p_{2}=\frac{1}{2}$. The QD expression depends on $\alpha_0$ and $\eta$ and  can be constructed straightforward by (\ref{eq:QD}). However, the expression is too lengthy to gain any insight from it. Instead of using the exact expression, we look for a simplified polynomial fitting, that allows us to have an easier understanding of the QD behavior for the photon added case. We use  the form,
\begin{equation}
D_{B_{\text{Aprox}}}(\rho^{AB}_{pa})=\sum_{i,j} c_{i,j}\alpha_0^i\eta^j,
\label{QDpaExpanded}
\end{equation}
that we have expanded up to the fourth power with the fitting coefficients $D_{B_{\text{Aprox}}}(\rho^{AB}_{pa})=0.0317084 \alpha _0^2 \eta ^3-0.398278 \alpha _0 \eta ^3-0.002986 \alpha _0^3 \eta ^2-0.0186853 \alpha _0^2 \eta ^2+0.617212 \alpha _0 \eta ^2+0.0557978 \alpha _0^2 \eta -0.7036 \alpha _0 \eta +0.000865144 \alpha _0^4-0.0192576 \alpha _0^3+0.133857 \alpha _0^2-0.280623 \alpha _0+0.519166 \eta ^4-0.466403 \eta ^3-0.308003 \eta ^2+1.113 \eta +0.140178$. This fitting is valid in the domain $\alpha_0=[0,10]$ and $\eta=[0.5,1]$. This fitting allows us to see the dominant correlations between the parameter $\alpha_0$ and $\eta$ that contribute to QD. It can be seen that for low $\alpha_0$, $\eta$ dominates; on the contrary, for large $\alpha_0$ the low coefficients accompanying the higher powers of $\alpha_0$ become significant, i.e., for $\alpha_0=0$ we can expect the same behavior as in DPC case, but as long as the average photon number $n_0=|\alpha_0|^2$ increases, the QD decreases.

\begin{figure}
\floatbox[{\capbeside\thisfloatsetup{capbesideposition={left,top},capbesidewidth=4cm}}]{figure}[\FBwidth]
{\caption{Quantum Discord from B to A with loss by scattering included for DPC (solid red line independent of $n_0$) and PAC (black lines) and its relation with $\Delta X_{\lambda_A=\pi/2}$ and $\eta$. In (a) QD vs. $\eta$, (b) $\Delta X_{\lambda_A=\pi/2}^2$ vs. $\eta$, (c) parametrization of QD vs. $\Delta X_{\lambda_A=\pi/2}^2$.}\label{fig:QDall}}
{\includegraphics[width=7cm]{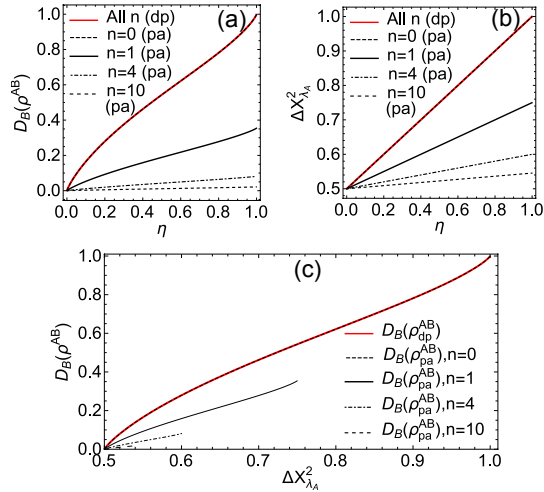}}
\end{figure}

Following the same procedure as before, Alice performs a large set of homodyne measurements on her subsystem trying to realize whether she can gain some information on the shared channel correlation by the quadrature variance. One can easily show that the variance squared is $\Delta X^{2}_{\lambda_A (\text{pa})}=\frac{\alpha _0^2 (2-\eta  \cos (2 \lambda ))+\alpha _0^4+\eta +1}{2 \left(\alpha _0^2+1\right){}^2}$, which shows the linear dependence on $\eta$ for low values of $\alpha_0$, as in the previous case. On the other hand, as $\alpha_0$ increases that dependency gets lost and $\alpha_0$ dominates. For a simpler understanding, if we select the phase $\lambda=\pi/2$ we  have the variance as $\Delta X^{2}_{\lambda_A (\text{pa})}=\frac{\text{$\alpha_0$}^2+\eta +1}{2 \text{$\alpha_0$}^2+2}$. Neither the exact nor the approximate expressions for the variance can be directly inverted to find an explicit QD dependence on  $\Delta X^{2}_{\lambda_A (\text{pa})}$, but it can be obtained numerically and plotted.
The variance of the state obtained by Alice is within the interval $[1/2,1]$. By acquiring the statistics of many quadratures, Alice can deduce the scattering noise $\eta$ and also a estimation of QD in the measurement scheme outlined above. The behavior of QD $D_B (\rho^{AB}_{\text{pd(pa)}})$ vs. ($\Delta X^2_{\lambda_A}, \eta)$ is shown in Fig. \ref{fig:QDall}. In \ref{fig:QDall}(a) we present  QD as a function of $\eta$, in Fig. \ref{fig:QDall} (b) the corresponding quadrature variance $\Delta X^2_{\lambda_A}$ for a given $\eta$, and in Fig. \ref{fig:QDall} (c) is shown the parametric plot of QD $D_B (\rho^{AB}_{\text{pd(pa)}})$ vs. ($\Delta X^2_{\lambda_A}, \eta)$. Only the red line corresponds to DPC, and in black lines is shown the effect on PAC for different values of photon number $n_0$. The red line is valid for DPC and for all values of the average photon number $n_0$ when only scattering is taken into account.  If one knows with certainty that the channel is DPC and that is only subjected to scattering, then the quadrature variance allows one to obtain the amount of QD and even the loss of the channel. Black lines show that for PAC, as one increases $n_0$ the slope of $\Delta X^2_{\lambda_A}$ vs. $\eta$ decreases, which is related to the amount of QD of the system. When $n_0=0$, $D_B(\rho^{AB}_{\text{pa}})$ coincides with $D_B(\rho^{AB}_{\text{dp}})$ but decreases as $n_0$ increases, along with the range of  $\Delta X^{2}_{\lambda_A=\pi/2 (\text{pa})}$ which also decreases. It is clear from the figure, that a measure of the variance can serve as a lower bound of QD even without the knowledge of $n_0$. Additionally, as the QD is related to $n_0$ and $\eta$, the knowledge of $n_0$ can allow to infer the scattering of the channel, this is clear by the dependence of (\ref{QDpaExpanded}) and $X^{2}_{\lambda_A (\text{pa})}$ with $\alpha_0$ and $\eta$.

\begin{figure}
\floatbox[{\capbeside\thisfloatsetup{capbesideposition={left,top},capbesidewidth=6cm}}]{figure}[\FBwidth]
{\caption{Quantum Discord from B to A for the pure channel dependent of $n_0$ for DPC (solid black line) and PAC (dashed black lines).}\label{fig:QDSinRuido}}
{\includegraphics[width=5cm]{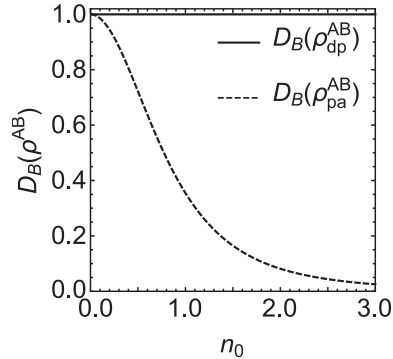}}
\end{figure}

Additionally, from Fig.\ref{fig:QDall}(a) the value of QD for $\eta=0$ can be obtained for both PAC and DPC, which is expected to be equivalent to entropy of entanglement as it is a pure state. The QD values for the extremal right top points of each line in Fig.\ref{fig:QDall}(c) in descending order are $D_B(\rho^{AB})=\{ 1,0.3545,0.0814,0.0214\}$ that correspond to the PAC case with $n_0=\{ 0,1,4,10 \}$ as in Fig.\ref{fig:QDall}(a). Evidently, for DPC $D_B(\rho^{AB})=1$ independent of $n_0$. These values of QD are shown in Fig. \ref{fig:QDSinRuido}, in the figure the QD for $\eta=1$ vs $n_0$ is plotted for DPC in solid line and for PAC in dashed line. From this figure we see that this behavior recovers the entropy of entanglement of each of the states that was shown in Fig. \ref{fig:entvsamid}, where also the AMID has the same qualitative behavior as QD. QD for DPC is independent of $n_0$ in the ideal channel, and QD for PAC is highly dependent of $n_0$ as expected from entropy and AMID.

\subsubsection{Loss by scattering and phase noise.}
Phase noise is added to the channel in order to complement the treatment of sources of loss, given the common nature of fluctuations in optical paths in any experiment. The channel considering both loss mechanisms is named $\rho_{\text{pd(pa)}}(\eta, \phi)$, as in (\ref{eq:phase1}). Unfortunately, the displacement with phase forbids the selection of an orthogonal basis as before. Instead, the QD is numerically analyzed by expanding (\ref{eq:phase1}) in the number basis, which can be easily accomplished in the standard manner  $\rho = \sum_{i,j,k,l}  \ket{ij}\bra{ij} \rho(\eta,\phi) \ket{kl}\bra{kl}$.

Following the same approach as in the previous section, QD is calculated evaluating (\ref{eq:QD}). Given the truncated expansion of the infinite matrices, the optimal angles are also numerically re-evaluated for each calculated point. The accuracy of QD obtained in the limit $\sigma \rightarrow 0$ proves to be equal as in previous calculations.

\begin{figure}
\floatbox[{\capbeside\thisfloatsetup{capbesideposition={left,top},capbesidewidth=4cm}}]{figure}[\FBwidth]
{\caption{QD from B to A for DPC vs phase variance $\sigma$ for different amounts of scattering $\eta$ and initial average photon number $n_0$.}\label{fig:QDartvssigma}}
{\includegraphics[width=7cm]{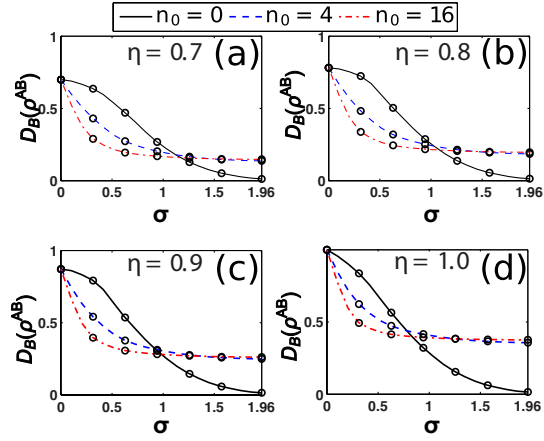}}
\end{figure}

\textbf{Displaced Photons Channel: DPC}. The new optimal measurement angles are $\theta_M^*$ and $\phi_M^*$ are no longer fixed, but sigma dependent, thereby they were found each time the QD was evaluated for any configuration of ($\eta, \sigma$). Also the quadrature variance with phase (QVP) fluctuation is obtained and is given by
\begin{equation}
\begin{aligned}
\Delta X_{\lambda_A (\text{dp})}^2(\lambda_A,\sigma)&=\frac{1}{2} e^{-2 \sigma ^2}
\Big(\alpha _0^2 \eta  \left(e^{\sigma ^2}-1\right) \\
&\quad \times\left(e^{\sigma ^2}-\cos \left(2 \lambda _A\right)\right) +(\eta +1) e^{2 \sigma ^2}\Big).
\end{aligned}
\label{eq:xlambda_phase_pd}
\end{equation}
 The angles $\lambda=0$ and $\lambda=\pi/2$ are selected to calculate QVP to see  the dependence of $\Delta X ^2_{\lambda_{A (\text{pd})}}$ with $\sigma$ in rads.
QD as a function of the phase variance $\sigma$ is shown in Fig. \ref{fig:QDartvssigma} for different values of scattering $\eta$ and average photon numbers $n_0$. From Fig. \ref{fig:QDartvssigma} (a) to (d), the scattering takes the values $\eta=[0.7,1.0]$ in steps of 0.1, additionally each panel shows three different average photon numbers $n_0$ for the initial state. Circle marks represent calculated points and the lines are  interpolations. The solid black line corresponds to $n_0=0$ which is equivalent to a traditional entangled channel of single photons of the form $\ket{01}+\ket{10}$ subjected to both loss mechanisms, the dashed blue and dotted red lines correspond to $n_0=4$ and $n_0=16$ photons respectively. The maximum value of obtainable QD is governed by the amount of scattering and is found for $\sigma=0$. This shows how QD for $\sigma=0$ increases up to its maximal attainable value of $D_B (\rho^{AB}_{\text{pd(pa)}})=1$ from (a) to (d). Two interesting facts arise from these plots: \textit{first}, the maximum QD belongs to the common $\sigma=0$ independent of $n_0$; \textit{second}, even when QD drops more rapidly for higher $n_0$ values for low values of $\sigma$, the channel QD, independently of photon number,  becomes resistant to phase fluctuations when $\sigma$ is large, as it tends to an asymptotic behavior higher than the QD of the channel compound only by the superposition of Fock states ($n_0=0$). In other words, when the hybrid continuous channel $\rho_{\text{pd}}$ is employed, a lower degradation of quantum correlations is experienced in a high phase fluctuating environment. All figures show similar qualitative behavior, where their differences rely on the maximum amount of QD that can be reached. For values of $\sigma$ greater than zero, QD becomes also dependent of $n_0$, as shown by the separation of the lines for $n_0=4$ and $n_0=16$ from the one for $n_0=0$ in the four figures \ref{fig:QDartvssigma}(a)-(d).

\begin{figure}
	\centering
	\includegraphics[width=0.55\textwidth]{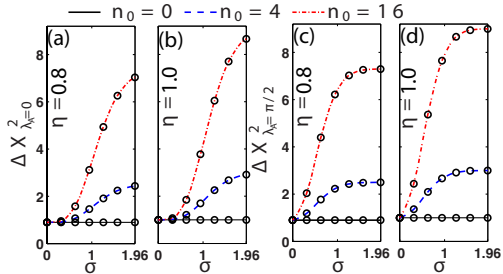}
	\caption{Quadrature variance squared (QVP) $\Delta{X_{\lambda_A}}^2$ vs. $\sigma$ for fixed values of $\eta$ for DPC. (a) $\eta=0.8, \lambda_A=0$, (b) $\eta=1.0, \lambda_A=0$, (c) $\eta=0.8, \lambda_A=\pi/2$, (d) $\eta=1.0, \lambda_A=\pi/2$.}
	\label{fig:DeltXVsSigmaArt}
\end{figure}

QVP squared $\Delta_{X_{\lambda_A=0(\pi/2)}}^2$ as a function of $\sigma$ for fixed values of $\eta$ is shown in Fig. \ref{fig:DeltXVsSigmaArt}. The quadratures were calculated by (\ref{eq:xlambda_phase_pd}).Fig. \ref{fig:DeltXVsSigmaArt} (a)-(b) with $\lambda_A=0$, and Fig. \ref{fig:DeltXVsSigmaArt} (c)-(d) with $\lambda_A=\pi/2$, for $\eta=\{0.8,1.0\}$. Solid black line corresponds to $n_0=0$, dashed blue and dotted-dashed red lines correspond to $n_0=4$ and $n_0=16$ respectively. The plots show that for a given value of $\sigma$, $\Delta_{X_{\lambda_A=0(\pi/2)}}^2 $ is not single-defined, but $\eta$ and $n_0$ dependent. However, the behavior of two parameters can be inferred from the other two. If the characteristics of the channel are known $(\eta, n_0)$,  then the phase variance $\sigma$ can be obtained by the behavior of QVP, as seen in (\ref{eq:xlambda_phase_pd}).

\begin{figure}
	\centering
	\includegraphics[width=0.85\textwidth]{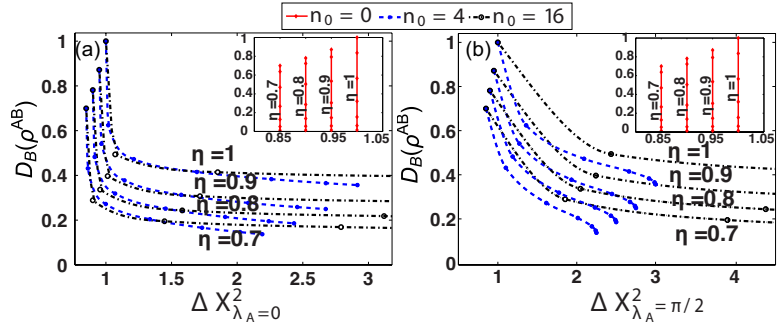}
	\caption{Parametric plot of QD from B to A for DPC vs: (a) $\Delta X_{\lambda_A=0}^2$, and (b)$\Delta X_{\lambda_A=\pi/2}^2$. QD is calculated for different values of $\eta$ and $n_0$. Marks represent $\sigma$ values in ascending order from top to bottom and left to right.}
	\label{fig:QDPDExp}
\end{figure}

Parameterized values of $\Delta X^2_{\lambda_{A (\text{pd})}}$ with $\sigma$ in the domain $[0,1.96]$ rad with its correspondent QD are shown in Fig. \ref{fig:QDPDExp}. These are a parametrization of the information shown in Figs. \ref{fig:QDartvssigma} and \ref{fig:DeltXVsSigmaArt} for $\lambda_A=0$ and $\lambda_A=\pi/2$. The plots are calculated for $\eta = \{0.7,1.0\}$  and different values of $n_0$. The average photon number goes from $n_0=0$ in the insets to $n_0=4$ in dashed blue lines and $n_0=16$ in black dashed lines. Fig. \ref{fig:QDPDExp}(a) shows the case for $\lambda_A=0$ and (b) $\lambda_A=\pi/2$. Each mark represents a pair $(D_B(\rho_{\text{dp}}^{AB}),\Delta X^2_{{\lambda_A}=\{0,\pi/2 \}})$ calculated as a function of $n_0, \eta, \sigma$. Following the marks from top to bottom and left to right each points' $\sigma$ is in increasing order (the lines are an interpolation).  The quasi-vertical lines shown in the inset figures correspond to $n_0=0$, this indicates that QVP is not a good indicator of QD for channels with $n_0=0$, because for tiny variations in QVP, high variations in QD are obtained. The behavior for $\sigma=0$ is recovered from top left points in each line. The relation of $\Delta X ^2_{\lambda_{A (\text{pd})}}$ when $\sigma$ increases is exhibited with the extension of the lines to the right, this occurs only for values of $n_0>0$. From both subplots the higher $\Delta X_{\lambda_A}^2$ the higher $\sigma$ is for a given $\eta$ (which could be easily determined) and therefore QD could be estimated as before.

\textbf{Photon Added Channel.} As in previous case, optimal measurement angles for QD were determined for every combination $(n_0,\eta,\sigma)$ of parameters analyzed. QVP at Alice's side is obtained as
\begin{equation}
	\Delta X^{2}_{\lambda_A (\text{pa})}(\lambda_A)= \frac{1}{2} \left(2 \eta +\frac{w_{\text{pa}}}{\left(\alpha _0^2+1\right){}^2}+1\right),
	\label{eq:xlambda_phase_pa}
\end{equation}
where $w_{\text{pa}}=\eta  e^{-2 \sigma ^2} (-2 \left(\alpha _0^3+2 \alpha _0\right){}^2 e^{\sigma ^2} \cos ^2(\lambda_A )+\left(\alpha _0^4+4 \alpha _0^2+3\right) \alpha _0^2 \cos (2 \lambda_A )+\left(\alpha _0^6+2 \alpha _0^4-1\right) e^{2 \sigma ^2})$ with both angles $\lambda_A=0,\pi/2$ explored as the previous case.

\begin{figure}
\floatbox[{\capbeside\thisfloatsetup{capbesideposition={left,top},capbesidewidth=4cm}}]{figure}[\FBwidth]
{\caption{QD from B to A for PAC vs phase variance $\sigma$ for different amounts of scattering $\eta$ and initial average photon number $n_0$.}\label{fig:QDprovssigma}}
{\includegraphics[width=7cm]{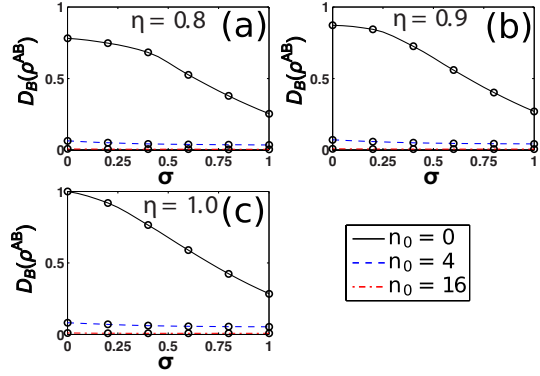}}
\end{figure}

In Fig. \ref{fig:QDprovssigma},  QD as a function of $\sigma$ is plotted for different values of $\eta$ (shown in panels from (a) to (c) with $\eta$ from 0.8, 0.9 and 1.0, respectively). Each panel shows three different $n_0$, where circle marks represent calculated points and the lines are interpolations. Solid black line is for the case $n_0=0$ and exactly the same behavior as in DPC is obtained. On the contrary, even small increments in $n_0$ drop QD close to zero as can be confirmed for the different $\eta$ explored.
In this case, the phase fluctuations destroys faster the correlations.

\begin{figure}
	\centering
	\includegraphics[width=0.55\textwidth]{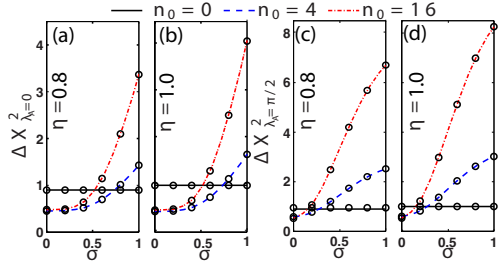}
	\caption{Quadrature variance squared (QVP) $\Delta_{X_{\lambda_A}}^2$ vs. $\sigma$ for fixed values of $\eta$ for PAC. (a) $\eta=0.8, \lambda_A=0$, (b) $\eta=1.0, \lambda_A=0$, (c) $\eta=0.8, \lambda_A=\pi/2$, (d) $\eta=1.0, \lambda_A=\pi/2$.}
	\label{fig:DeltXVsSigmaPro}
\end{figure}

QVP squared $\Delta{X^2_{\lambda_A=0(\pi/2)}}$ as a function of $\sigma$ for fixed values of $\eta$ are shown in Fig. \ref{fig:DeltXVsSigmaPro}. The quadratures were calculated by (\ref{eq:xlambda_phase_pa}). Fig. \ref{fig:DeltXVsSigmaPro} (a)-(b) for $\lambda_A=0$, and Fig. \ref{fig:DeltXVsSigmaPro}  (c)-(d) for $\lambda_A=\pi/2$, with $\eta=\{0.8,1.0\}$. Solid black line corresponds to $n_0=0$, dashed blue and dotted-dashed black lines correspond to $n_0=4$ and $n_0=16$ respectively. As in previous case, the plots show that for a given value of $\sigma$, $\Delta{X^2_{\lambda_A=0(\pi/2)}} $ is not single-defined, but $\eta$ and $n_0$ dependent; additionally, for $n_0 \ne 0$ QVP starts from lower values than $\Delta_{X_{\lambda_A}}^2=1$, what in this case, indicates that the state behaves more like Gaussian, and closer to the expected vacuum fluctuation, in other words, its a signature of lower correlations between quadratures. As in Fig. \ref{fig:DeltXVsSigmaArt}, the behavior of two parameters can be inferred from the other two.

\begin{figure}
	\centering
	\includegraphics[width=0.85\textwidth]{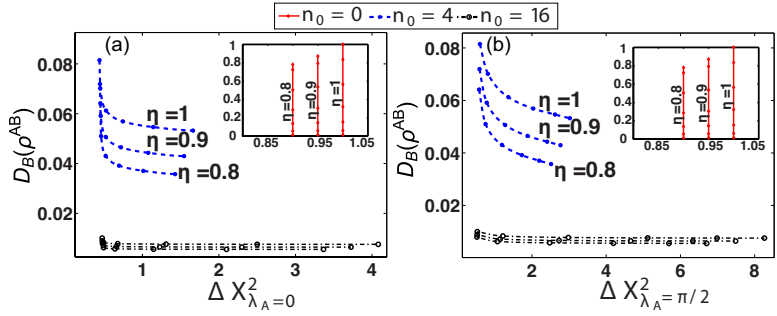}
	\caption{Parametric plot of QD from B to A for PAC vs: (a) $\Delta X_{\lambda_A=0}^2$, and (b)$\Delta X_{\lambda_A=\pi/2}^2$. QD is calculated for different values of $\eta$ and $n_0$. Marks represent $\sigma$ values in ascending order from top to bottom and left to right.}
	\label{fig:QDPAExp}
\end{figure}

In Fig. \ref{fig:QDPAExp} a parametric plot of $\Delta X_{\lambda_A (pa)}^2$ versus  QD is shown with $\sigma$ in the domain of $[0,1]$rad in a similar way as in the DPC. This results as a parametrization of the information contained in Fig. \ref{fig:QDprovssigma} and \ref{fig:DeltXVsSigmaPro} for $\lambda_A=\{0,\pi/2\}$ shown in Fig. \ref{fig:QDPAExp}(a) and (b) respectively.  The plots are calculated for $\eta = \{0.8,0.9,1.0\}$ and different values of $n_0$. The average photon number goes from $n_0=0$ in the insets to $n_0=4$ in dashed blue lines and $n_0=16$ in black dashed-dotted lines. Same behavior is obtained as in DPC for $n_0=0$ with the continuous red quasivertical lines where QD decreases as $\sigma$ grows, while all $n_0>0$ scenarios have almost zero discord. A QVP measure for the PAC result of low practical use to infer any characteristic correlation
of the channel under this configuration as can be seen by the long almost horizontal lines (dashed blue and dotted black) for both figures. This indicates that for PAC is hard to set apart different amounts of QD for any given QVP, mainly because QD decays to values close to zero rapidly.

\section{Conclusions}
In this work we have characterized two bipartite states formed by linear combination of SPACS and DFS as possible models for communication channels, DPC and PAC respectively, subjected to scattering and phase noise as loss mechanisms. In the estimation of quantum correlations of the channel we started by JQP distributions, where this estimator gave information about the correlations of the channel that could be exploited in a QCP. Quadrature MID characterization of the pure channel served as a confirmation of applicability of quadrature measurements as an entropy estimator since quadratures are easily acquired experimentally.  The quantum discord was used as a more quantitative estimator  and  was calculated for both channels for different combinations of $(n_0, \eta, \sigma)$, and the dependency of QD as a function QVP was explored as a means to determine the channel correlation with a local measurement. It was found that QD for DPC exhibits more robustness to high phase fluctuations when $n_0>0$, because the minimum QD obtainable with an increasing $\sigma$ increases compared to the case $n_0=0$; this is not valid for PAC where any increase in $n_0>0$ drops the QD to a value close to zero. The response of QD vs QVP was also obtained as a parametrization of QD with $\Delta X_{\lambda_A}^2$, where QVP serves to characterize QD for DPC but not for PAC. In summary, we have presented a study of the non-classical properties of a bipartite channel that could be useful in a future work applied  to a quantum communication protocol.

\section{Acknowledgements}
The first author acknowledges receipt of a PhD grant number 331668 from CONACYT.

\appendix

\section{\label{app1}MID calculation details}

All details of MID calculation shown in Section \ref{MIDpure} are addressed in this section, as long as the expressions used to obtain AMID. To obtain (\ref{eq:MIDdef}) and (\ref{eq:AMIDdef}) we need mutual information for the pure system and for the projected one. Pure system entropy is easily computed with the Von Neumann formula in the traditional form as in (\ref{eq:entVNrhoA_PDPA}). And for the entropies of the quadrature projected versions equations (\ref{eq:MID5}) and (\ref{eq:MID6}) are used. One may note that these are CV entropies and therefore are not invariant to change of basis \cite{Ash65}, for this reason the entropies in the quadrature basis are not the same as in the number basis. The $\lambda$-dependent entropies are shown for DPC and PAC cases respectively. For any reduced subsystem of the DPC case is constant with the value
		\begin{equation}
			S(\rho^{A(B)}_{\text{pd}}) (\lambda_{A(B)}) \approx 2.00208,
			\label{eq:Spd1}
		\end{equation}
while the entropy for the complete DPC system is				
		\begin{equation}
		\tiny
			\begin{aligned}
				S&(\rho^{AB}_{\text{pd}})  (\lambda_A,\lambda_B)= \int_{-\infty}^{\infty} \text{dX}_{\lambda_A} \int_{-\infty}^{\infty} \text{dX}_{\lambda_B} \\
				& \quad  \frac{\exp \left(-\frac{1}{2} \alpha _0^2 \left(\cos \left(2 \lambda _A\right)+\cos \left(2 \lambda _B\right)+2\right)+2 \alpha _0 \left(X_{\lambda _A} \cos \left(\lambda _A\right)+X_{\lambda _B} \cos \left(\lambda _B\right)\right)-X_{\lambda _A}^2-X_{\lambda _B}^2\right)}{\pi  \ln (16)}\\
				&\quad\times \Big( \left| 2 e^{i \lambda _A} X_{\lambda _A}-\alpha _0\right| {}^2+2 \Re\left(e^{i \lambda _B-2 i \lambda _A} \left(-\alpha _0+2 e^{i \lambda _A} X_{\lambda _A}\right) \left(-2 X_{\lambda _B}+\alpha _0 e^{i \lambda _B}\right)\right)\\
				&\quad + \left| 2 e^{i \lambda _B} X_{\lambda _B}-\alpha _0\right| {}^2 \Big) \times \Big( -\ln \big(\left| 2 e^{i \lambda _A} X_{\lambda _A}-\alpha _0\right| {}^2+2 \Re\Big(e^{i \lambda _B-2 i \lambda _A} \left(-\alpha _0+2 e^{i \lambda _A} X_{\lambda _A}\right) \\
				& \quad \times \left(-2 X_{\lambda _B}+\alpha _0 e^{i \lambda _B}\right)\Big)+\left| 2 e^{i \lambda _B} X_{\lambda _2}-\alpha _0\right| {}^2\big)\\
				&\quad+\frac{1}{2} \alpha _0^2 \left(\cos \left(2 \lambda _A\right)+\cos \left(2 \lambda _B\right)+2\right)-2 \alpha _0 \left(X_{\lambda _1} \cos \left(\lambda _A\right)+X_{\lambda _2} \cos \left(\lambda _B\right)\right)+X_{\lambda _1}^2+X_{\lambda _2}^2+\log (4)+\ln (\pi ) \Big).
			\end{aligned}
			\label{{eq:Spd2}}
		\end{equation}	
On the other hand, both entropies for PAC are $\lambda$-dependent, as shown in the entropy for the reduced subsystems	
		\begin{equation}
		\tiny
		\begin{aligned}
		S & (\rho^{A(B)}_{\text{pa}}) (\lambda_{A(B)})=\\
		&\quad -\int\limits_{-\infty}^{\infty} \text{dX}_{\lambda_{A(B)}}  \left(-2 \alpha_0^2 \cos (2 \lambda_{A(B)})+\alpha_0^2+\left| 2 e^{i \lambda_{A(B)}} X_{\lambda_{A(B)}}-\alpha_0\right|^2+4 \alpha_0 X_{\lambda_{A(B)}} \cos (\lambda_{A(B)})+2\right) \\
		&\quad \times \ln \left(\frac{-2 \alpha_0^2 \cos (2 \lambda_{A(B)})+\alpha_0^2+\left| 2 e^{i \lambda_{A(B)}} X_{\lambda_{A(B)}}-\alpha_0\right|^2+4 \alpha_0 X_{\lambda_{A(B)}} \cos (\lambda_{A(B)})+2}{\xi_1}\right)/(\xi_1 \ln 2),
		\end{aligned}
		\label{eq:Spa1}
		\end{equation}
with $\xi_1=4 \sqrt{\pi } \left(\alpha_0^2+1\right) (\sinh ((X_{\lambda_{A(B)}}-\alpha_0 \cos (\lambda_{A(B)}))^2) + \cosh \left((X_{\lambda_{A(B)}}-\alpha_0 \cos (\lambda_{A(B)}))^2\right))$. And for the complete PAC system,		
		\begin{equation}
		\tiny
		\begin{aligned}
		S&(\rho^{AB}_{\text{pa}})  (\lambda_A,\lambda_B)= -  \int_{-\infty}^{\infty} \text{dX}_{\lambda_A} \int_{-\infty}^{\infty} \text{dX}_{\lambda_B} \\
		& \frac{\exp \left(-\frac{1}{2} \alpha _0^2 \left(\cos \left(2 \lambda _A\right)+\cos \left(2 \lambda _B\right)+2\right)+2 \alpha _0 \left(X_{\lambda _1} \cos \left(\lambda _A\right)+X_{\lambda _2} \cos \left(\lambda _B\right)\right)-X_{\lambda _1}^2-X_{\lambda _2}^2\right)}{4 \pi  \left(\alpha _0^2+1\right) \ln (2)}\\
		&\times \ln \Big[\Big(\left| 2 e^{i \lambda _A} X_{\lambda _1}-\alpha _0\right| {}^2-2 \Re\left(e^{i \left(\lambda _A-2 \lambda _B\right)} \left(-2 X_{\lambda _1}+\alpha _0 e^{i \lambda _A}\right) \left(-\alpha _0+2 e^{i \lambda _B} X_{\lambda _2}\right)\right) \\
		&+\left| 2 e^{i \lambda _B} X_{\lambda _2}-\alpha _0\right| {}^2\Big)  \times \exp \Big(-\frac{1}{2} \alpha _0^2 \left(\cos \left(2 \lambda _A\right)+\cos \left(2 \lambda _B\right)+2\right)+2 \alpha _0 \left(X_{\lambda _1} \cos \left(\lambda _A\right)+X_{\lambda _2} \cos \left(\lambda _B\right)\right)\\
		& -X_{\lambda _1}^2-X_{\lambda _2}^2\Big) / 4 \pi  \left(\alpha _0^2+1\right)     \Big] 	\Big(\left| 2 e^{i \lambda _A} X_{\lambda _1}-\alpha _0\right| {}^2\\
		&-2 \Re\left(e^{i \left(\lambda _A-2 \lambda _B\right)} \left(-2 X_{\lambda _1}+\alpha _0 e^{i \lambda _A}\right) \left(-\alpha _0+2 e^{i \lambda _B} X_{\lambda _2}\right)\right)+\left| 2 e^{i \lambda _B} X_{\lambda _2}-\alpha _0\right| {}^2\Big),
		\end{aligned}
		\label{eq:Spa2}
		\end{equation}

The expression for (\ref{eq:Spd1}) has not been included explicitly, but only its approximate value $\forall \{\alpha_0,\lambda_A,\lambda_B\}$.

MID is evaluated using (\ref{eq:MIDdef}) through previous expressions from (\ref{eq:Spd1}) to (\ref{eq:Spa2}) and is plotted in Fig. \ref{fig:MIDmap} from where one can select the angles that maximize MID. By following this inspection, a set of angles for $\lambda_A$ and $\lambda_B$ are selected to simplify $\lambda$-dependent entropy expressions. MID has minimum for the DPC case when $\lambda_A=\lambda_B$, i.e., one minimum for the complete DPC case is obtained as,
		\begin{equation}
		\tiny
		\begin{aligned}
		S (\rho^{AB}_{\text{pd}}) &  (\lambda_A=0,\lambda_B=0)=-\frac{2}{\pi  \ln (4)} \int \limits_{-\infty }^{\infty } \text{dX}_{\lambda_A} \int \limits_{-\infty }^{\infty }  \text{dX}_{\lambda_B} (X_{\lambda_A}-X_{\lambda_B})^2 \\
		&\quad \times e^{-2 \alpha_0^2-X_{\lambda_A}^2+2 \alpha_0 (X_{\lambda_A}+X_{\lambda_B})-X_{\lambda_B}^2} \\
		&\quad \times \ln \left(\frac{(X_{\lambda_A}-X_{\lambda_B})^2 e^{-2 \alpha_0^2-X_{\lambda_A}^2+2 \alpha_0 (X_{\lambda_A}+X_{\lambda_B})-X_{\lambda_B}^2}}{\pi }\right).
		\end{aligned}
		\label{eq:MIDO2PD}
		\end{equation}

For PAC case one of the minimums is found for $\lambda_A=\lambda_B=0$, what produces the maximum entropy for one subsystem as
		\begin{equation}
		\tiny
		\begin{aligned}
			S(\rho^{A(B)}_{\text{pa}}) & (\lambda_{A(B)}=0)= -\int\limits_{-\infty }^{\infty } \text{dX}_{\lambda_{A(B)}}\\
			&\quad\times \frac{\left(2 X_{\lambda_{A(B)}}^2+1\right) e^{-(X_{\lambda_{A(B)}}-\alpha_0)^2} \left(\ln \left(\frac{2 X_{\lambda_{A(B)}}^2+1}{2 \alpha_0^2+2}\right)-(X_{\lambda_{A(B)}}-\alpha_0)^2-\frac{\ln (\pi )}{2}\right)}{\sqrt{\pi } \left(\alpha_0^2+1\right) \ln (4)},
			\end{aligned}
			\label{eq:MIDO1PA}
		\end{equation}
while one of the minimal entropy points for complete PAC system is obtained as		
		\begin{equation}
		\tiny
			\begin{aligned}
				S(\rho^{AB}_{\text{pa}})&  (\lambda_A=0,\lambda_B=0)=-\frac{1}{\pi  \left(\text{$\alpha $0}^2+1\right) \ln (2)}\int \limits_{-\infty }^{\infty }\text{dX}_{\lambda_A}\int \limits_{-\infty }^{\infty }  \text{dX}_{\lambda_B} e^{-2 \alpha_0^2-X_{\lambda_A}^2+2 \alpha_0 X_{\lambda_A}-X_{\lambda_B}^2+2 \alpha_0 X_{\lambda_B}}\\
				&\quad \times (-\alpha_0+X_{\lambda_A}+X_{\lambda_B})^2  \ln \left(\frac{e^{-2 \alpha_0^2-X_{\lambda_A}^2+2 \alpha_0 X_{\lambda_A}-X_{\lambda_B}^2+2 \alpha_0 X_{\lambda_B}} (-\alpha_0+X_{\lambda_A}+X_{\lambda_B})^2}{\pi  \left(\alpha_0^2+1\right)}\right).
			\end{aligned}
			\label{eq:MIDO2PA}
		\end{equation}
		\label{eq:Slalb_pa_simp}


\begin{thebibliography}{10}
\expandafter\ifx\csname url\endcsname\relax
  \def\url#1{\texttt{#1}}\fi
\expandafter\ifx\csname urlprefix\endcsname\relax\def\urlprefix{URL }\fi
\providecommand{\bibinfo}[2]{#2}
\providecommand{\eprint}[2][]{\url{#2}}

\bibitem{Aolita2015}
\bibinfo{author}{Aolita, L.}, \bibinfo{author}{de~Melo, F.} \&
  \bibinfo{author}{Davidovich, L.}
\newblock \bibinfo{title}{{Open-system dynamics of entanglement:a key issues
  review}}.
\newblock \emph{\bibinfo{journal}{Rep. Prog. Phys.}}
  \textbf{\bibinfo{volume}{78}}, \bibinfo{pages}{042001}
  (\bibinfo{year}{2015}).

\bibitem{Jozsa2003}
\bibinfo{author}{Jozsa, R.} \& \bibinfo{author}{Linden, N.}
\newblock \bibinfo{title}{{On the role of entanglement in quantum-computational
  speed-up}}.
\newblock \emph{\bibinfo{journal}{Proceedings of the Royal Society A:
  Mathematical, Physical and Engineering Sciences}}
  \textbf{\bibinfo{volume}{459}}, \bibinfo{pages}{2011--2032}
  (\bibinfo{year}{2003}).

\bibitem{Driessen2013}
\bibinfo{author}{Driessen, E. F.~C.}
\newblock \bibinfo{title}{{Single-photon detectors: Fast and efficient}}.
\newblock \emph{\bibinfo{journal}{Nature Photonics}}
  \textbf{\bibinfo{volume}{7}}, \bibinfo{pages}{168--169}
  (\bibinfo{year}{2013}).

\bibitem{Mirza2016}
\bibinfo{author}{Mirza, I.~M.} \& \bibinfo{author}{Schotland, J.~C.}
\newblock \bibinfo{title}{{Two-photon entanglement in multiqubit
  bidirectional-waveguide QED}}.
\newblock \emph{\bibinfo{journal}{Physical Review A}}
  \textbf{\bibinfo{volume}{94}} (\bibinfo{year}{2016}).
\newblock \eprint{1604.03652}.

\bibitem{Mirza2015}
\bibinfo{author}{Mirza, I.~M.}
\newblock \bibinfo{title}{{Bi- and uni-photon entanglement in two-way cascaded
  fiber-coupled atom-cavity systems}}.
\newblock \emph{\bibinfo{journal}{Physics Letters, Section A: General, Atomic
  and Solid State Physics}} \textbf{\bibinfo{volume}{379}},
  \bibinfo{pages}{1643--1648} (\bibinfo{year}{2015}).

\bibitem{Mirza2016a}
\bibinfo{author}{Mirza, I.~M.} \& \bibinfo{author}{Schotland, J.~C.}
\newblock \bibinfo{title}{{Multi-qubit entanglement in bi-directional chiral
  waveguide QED}}.
\newblock \emph{\bibinfo{journal}{Physical Review A}}
  \textbf{\bibinfo{volume}{94}}, \bibinfo{pages}{012302}
  (\bibinfo{year}{2016}).

\bibitem{Obrien2010}
\bibinfo{author}{O'Brien, J.~L.}, \bibinfo{author}{Furusawa, A.} \&
  \bibinfo{author}{Vu{\v{c}}kovi{\'{c}}, J.}
\newblock \bibinfo{title}{{Photonic quantum technologies}}.
\newblock \emph{\bibinfo{journal}{Nature Photonics}}
  \textbf{\bibinfo{volume}{3}}, \bibinfo{pages}{687--695}
  (\bibinfo{year}{2010}).

\bibitem{Togan2010}
\bibinfo{author}{Togan, E.} \emph{et~al.}
\newblock \bibinfo{title}{{Quantum entanglement between an optical photon and a
  solid-state spin qubit.}}
\newblock \emph{\bibinfo{journal}{Nature}} \textbf{\bibinfo{volume}{466}},
  \bibinfo{pages}{730--734} (\bibinfo{year}{2010}).

\bibitem{Volz2006}
\bibinfo{author}{Volz, J.} \emph{et~al.}
\newblock \bibinfo{title}{{Observation of entanglement of a single photon with
  a trapped atom}}.
\newblock \emph{\bibinfo{journal}{Physical Review Letters}}
  \textbf{\bibinfo{volume}{96}} (\bibinfo{year}{2006}).

\bibitem{Stute2013}
\bibinfo{author}{Stute, a.} \emph{et~al.}
\newblock \bibinfo{title}{{Tunable ion-photon entanglement in an optical
  cavity}}.
\newblock \emph{\bibinfo{journal}{Nature}} \textbf{\bibinfo{volume}{485}},
  \bibinfo{pages}{482--485} (\bibinfo{year}{2013}).

\bibitem{Braunstein2005}
\bibinfo{author}{Braunstein, S.~L.}
\newblock \bibinfo{title}{{Quantum information with continuous variables}}.
\newblock \emph{\bibinfo{journal}{Reviews of Modern Physics}}
  \textbf{\bibinfo{volume}{77}}, \bibinfo{pages}{513--577}
  (\bibinfo{year}{2005}).

\bibitem{Masada15}
\bibinfo{author}{Masada, G.} \emph{et~al.}
\newblock \bibinfo{title}{{Continuous-variable entanglement on a chip}}.
\newblock \emph{\bibinfo{journal}{Nature Photonics}}
  \textbf{\bibinfo{volume}{9}}, \bibinfo{pages}{316----319}
  (\bibinfo{year}{2015}).

\bibitem{Takeda2013a}
\bibinfo{author}{Takeda, S.}, \bibinfo{author}{Mizuta, T.},
  \bibinfo{author}{Fuwa, M.}, \bibinfo{author}{van Loock, P.} \&
  \bibinfo{author}{Furusawa, A.}
\newblock \bibinfo{title}{{Deterministic quantum teleportation of photonic
  quantum bits by a hybrid technique.}}
\newblock \emph{\bibinfo{journal}{Nature}} \textbf{\bibinfo{volume}{500}},
  \bibinfo{pages}{315--8} (\bibinfo{year}{2013}).

\bibitem{Sherson2006}
\bibinfo{author}{Sherson, J.~F.} \emph{et~al.}
\newblock \bibinfo{title}{{Quantum teleportation between light and matter.}}
\newblock \emph{\bibinfo{journal}{Nature}} \textbf{\bibinfo{volume}{443}},
  \bibinfo{pages}{557--560} (\bibinfo{year}{2006}).

\bibitem{Gu2012}
\bibinfo{author}{Gu, M.} \emph{et~al.}
\newblock \bibinfo{title}{{Observing the operational significance of
  discord consumption}}.
\newblock \emph{\bibinfo{journal}{Nature Physics}}
  \textbf{\bibinfo{volume}{8}}, \bibinfo{pages}{671--675}
  (\bibinfo{year}{2012}).

\bibitem{Modi2012}
\bibinfo{author}{Modi, K.}, \bibinfo{author}{Brodutch, A.},
  \bibinfo{author}{Cable, H.}, \bibinfo{author}{Paterek, T.} \&
  \bibinfo{author}{Vedral, V.}
\newblock \bibinfo{title}{{The classical-quantum boundary for correlations:
  Discord and related measures}}.
\newblock \emph{\bibinfo{journal}{Reviews of Modern Physics}}
  \textbf{\bibinfo{volume}{84}}, \bibinfo{pages}{1655--1707}
  (\bibinfo{year}{2012}).

\bibitem{Laflamme2001}
\bibinfo{author}{Laflamme, R.}, \bibinfo{author}{Cory, D.~G.},
  \bibinfo{author}{Negrevergne, C.} \& \bibinfo{author}{Viola, L.}
\newblock \bibinfo{title}{{NMR Quantum Information Processing and
  Entanglement}}.
\newblock \emph{\bibinfo{journal}{Quant. Inf. Comput.}}
  \textbf{\bibinfo{volume}{2}}, \bibinfo{pages}{166--176}
  (\bibinfo{year}{2001}).

\bibitem{Dakic2012a}
\bibinfo{author}{Daki{\'{c}}, B.} \emph{et~al.}
\newblock \bibinfo{title}{{Quantum discord as resource for remote state
  preparation}}.
\newblock \emph{\bibinfo{journal}{Nature Physics}}
  \textbf{\bibinfo{volume}{8}}, \bibinfo{pages}{666--670}
  (\bibinfo{year}{2012}).

\bibitem{Lanyon2013}
\bibinfo{author}{Lanyon, B.~P.} \emph{et~al.}
\newblock \bibinfo{title}{{Experimental generation of quantum discord via noisy
  processes}}.
\newblock \emph{\bibinfo{journal}{Physical Review Letters}}
  \textbf{\bibinfo{volume}{111}} (\bibinfo{year}{2013}).

\bibitem{Benedetti2013}
\bibinfo{author}{Benedetti, C.}, \bibinfo{author}{Shurupov, A.~P.},
  \bibinfo{author}{Paris, M. G.~A.}, \bibinfo{author}{Brida, G.} \&
  \bibinfo{author}{Genovese, M.}
\newblock \bibinfo{title}{{Experimental estimation of quantum discord for a
  polarization qubit and the use of fidelity to assess quantum correlations}}.
\newblock \emph{\bibinfo{journal}{Physical Review A - Atomic, Molecular, and
  Optical Physics}} \textbf{\bibinfo{volume}{87}} (\bibinfo{year}{2013}).

\bibitem{Fanchini2010}
\bibinfo{author}{Fanchini, F.~F.}, \bibinfo{author}{Werlang, T.},
  \bibinfo{author}{Brasil, C.~A.}, \bibinfo{author}{Arruda, L. G.~E.} \&
  \bibinfo{author}{Caldeira, A.~O.}
\newblock \bibinfo{title}{{Non-Markovian dynamics of quantum discord}}.
\newblock \emph{\bibinfo{journal}{Physical Review A}}
  \textbf{\bibinfo{volume}{81}}, \bibinfo{pages}{052107}
  (\bibinfo{year}{2010}).

\bibitem{FernandesFanchini2017}
\bibinfo{author}{{Fernandes Fanchini}, F.}, \bibinfo{author}{{Soares Pinto}, D.
  d.~O.} \& \bibinfo{author}{Adesso, G.}
\newblock \emph{\bibinfo{title}{{Lectures on General Quantum Correlations and
  their Applications}}} (\bibinfo{publisher}{Springer International
  Publishing}, \bibinfo{year}{2017}).

\bibitem{Xu2017}
\bibinfo{author}{Xu, J.-S.}, \bibinfo{author}{Li, C.-F.} \&
  \bibinfo{author}{Guo, G.-C.}
\newblock \emph{\bibinfo{title}{{Experimental Investigation of the Dynamics of
  Quantum Discord in Optical Systems}}}, \bibinfo{pages}{473--484}
  (\bibinfo{publisher}{Springer International Publishing},
  \bibinfo{address}{Cham}, \bibinfo{year}{2017}).

\bibitem{Hosseini2014}
\bibinfo{author}{Hosseini, S.} \emph{et~al.}
\newblock \bibinfo{title}{{Experimental verification of quantum discord in
  continuous-variable states}}.
\newblock \emph{\bibinfo{journal}{Journal of Physics B: Atomic, Molecular and
  Optical Physics}} \textbf{\bibinfo{volume}{47}}, \bibinfo{pages}{025503}
  (\bibinfo{year}{2014}).

\bibitem{Hosseini2014a}
\bibinfo{author}{Hosseini, S.} \emph{et~al.}
\newblock \bibinfo{title}{{Experimental verification of quantum discord in
  continuous-variable states and operational significance of discord
  consumption}}.
\newblock \emph{\bibinfo{journal}{Conference on Lasers and Electro-Optics
  Europe - Technical Digest}} \textbf{\bibinfo{volume}{2014-Janua}},
  \bibinfo{pages}{3--4} (\bibinfo{year}{2014}).

\bibitem{Ferraro2010}
\bibinfo{author}{Ferraro, A.}, \bibinfo{author}{Aolita, L.},
  \bibinfo{author}{Cavalcanti, D.}, \bibinfo{author}{Cucchietti, F.~M.} \&
  \bibinfo{author}{Ac{\'{i}}n, A.}
\newblock \bibinfo{title}{{Almost all quantum states have nonclassical
  correlations}}.
\newblock \emph{\bibinfo{journal}{Physical Review A - Atomic, Molecular, and
  Optical Physics}} \textbf{\bibinfo{volume}{81}} (\bibinfo{year}{2010}).

\bibitem{Jeong08a}
\bibinfo{author}{Jeong, H.} \& \bibinfo{author}{Kim, M.~S.}
\newblock \bibinfo{title}{{Efficient quantum computation using coherent
  states}}.
\newblock \emph{\bibinfo{journal}{arXiv:quant-ph}}
  \textbf{\bibinfo{volume}{v2}}, \bibinfo{pages}{1--6} (\bibinfo{year}{2001}).

\bibitem{Daoud2012}
\bibinfo{author}{Daoud, M.} \& \bibinfo{author}{Laamara, R.~A.}
\newblock \bibinfo{title}{{Quantum discord of Bell cat states under amplitude
  damping}}.
\newblock \emph{\bibinfo{journal}{Journal of Physics A: Mathematical and
  Theoretical}} \textbf{\bibinfo{volume}{45}}, \bibinfo{pages}{325302}
  (\bibinfo{year}{2012}).

\bibitem{Agarwal1991}
\bibinfo{author}{Agarwal, G.} \& \bibinfo{author}{Tara, K.}
\newblock \bibinfo{title}{{Nonclassical properties of states generated by the
  excitations on a coherent state}}.
\newblock \emph{\bibinfo{journal}{Physical Review A}}
  \textbf{\bibinfo{volume}{43}}, \bibinfo{pages}{492--497}
  (\bibinfo{year}{1991}).

\bibitem{Zavatta2004}
\bibinfo{author}{Zavatta, A.}, \bibinfo{author}{Viciani, S.} \&
  \bibinfo{author}{Bellini, M.}
\newblock \bibinfo{title}{{Quantum-to-Classical Transition with
  Single-Photon-Added Coherent States of Light}}.
\newblock \emph{\bibinfo{journal}{Science}} \textbf{\bibinfo{volume}{306}},
  \bibinfo{pages}{660--662} (\bibinfo{year}{2004}).

\bibitem{Kenfack2004}
\bibinfo{author}{Kenfack, A.} \& \bibinfo{author}{Yczkowski, K.}
\newblock \bibinfo{title}{{Negativity of the Wigner function as an indicator of
  non-classicality}}.
\newblock \emph{\bibinfo{journal}{Journal of Optics B: Quantum and
  Semiclassical Optics}} \textbf{\bibinfo{volume}{6}},
  \bibinfo{pages}{396--404} (\bibinfo{year}{2004}).

\bibitem{Kim2002}
\bibinfo{author}{Kim, M.~S.}, \bibinfo{author}{Son, W.},
  \bibinfo{author}{Bu{\v{z}}ek, V.} \& \bibinfo{author}{Knight, P.~L.}
\newblock \bibinfo{title}{{Entanglement by a beam splitter: Nonclassicality as
  a prerequisite for entanglement}}.
\newblock \emph{\bibinfo{journal}{Physical Review A}}
  \textbf{\bibinfo{volume}{65}}, \bibinfo{pages}{032323}
  (\bibinfo{year}{2002}).

\bibitem{Sekatski2012a}
\bibinfo{author}{Sekatski, P.} \emph{et~al.}
\newblock \bibinfo{title}{{Proposal for exploring macroscopic entanglement with
  a single photon and coherent states}}.
\newblock \emph{\bibinfo{journal}{Physical Review A}}
  \textbf{\bibinfo{volume}{86}}, \bibinfo{pages}{060301}
  (\bibinfo{year}{2012}).

\bibitem{Wang2015}
\bibinfo{author}{Wang, S.}, \bibinfo{author}{Hou, L.-L.},
  \bibinfo{author}{Chen, X.-F.} \& \bibinfo{author}{Xu, X.-F.}
\newblock \bibinfo{title}{{Continuous-variable quantum teleportation with
  non-Gaussian entangled states generated via multiple-photon subtraction and
  addition}}.
\newblock \emph{\bibinfo{journal}{Physical Review A}}
  \textbf{\bibinfo{volume}{91}}, \bibinfo{pages}{063832}
  (\bibinfo{year}{2015}).

\bibitem{Jeong2014}
\bibinfo{author}{Jeong, H.} \emph{et~al.}
\newblock \bibinfo{title}{{Generation of hybrid entanglement of light}}.
\newblock \emph{\bibinfo{journal}{Nature Photonics}}
  \textbf{\bibinfo{volume}{8}}, \bibinfo{pages}{564--569}
  (\bibinfo{year}{2014}).

\bibitem{Kwon2015}
\bibinfo{author}{Kwon, H.} \& \bibinfo{author}{Jeong, H.}
\newblock \bibinfo{title}{{Generation of hybrid entanglement between a
  single-photon polarization qubit and a coherent state}}.
\newblock \emph{\bibinfo{journal}{Physical Review A - Atomic, Molecular, and
  Optical Physics}} \textbf{\bibinfo{volume}{91}} (\bibinfo{year}{2015}).

\bibitem{Silva2006}
\bibinfo{author}{Silva, M. B. C.~E.}, \bibinfo{author}{Xu, Q.},
  \bibinfo{author}{Agnolini, S.}, \bibinfo{author}{Gallion, P.} \&
  \bibinfo{author}{Mendieta, F.~J.}
\newblock \bibinfo{title}{{Homodyne QPSK Detection for Quantum Key
  Distribution}}.
\newblock In \emph{\bibinfo{booktitle}{Optical Amplifiers and Their
  Applications/Coherent Optical Technologies and Applications}},
  \bibinfo{number}{2}, \bibinfo{pages}{CFA2} (\bibinfo{publisher}{Optical
  Society of America}, \bibinfo{year}{2006}).

\bibitem{Chuan2010}
\bibinfo{author}{Chuan, W.}, \bibinfo{author}{Wan-Ying, W.},
  \bibinfo{author}{Qing, A.} \& \bibinfo{author}{Gui-Lu, L.}
\newblock \bibinfo{title}{{Deterministic Quantum Key Distribution with Pulsed
  Homodyne Detection}}.
\newblock \emph{\bibinfo{journal}{Communications in Theoretical Physics}}
  \textbf{\bibinfo{volume}{53}}, \bibinfo{pages}{67--70}
  (\bibinfo{year}{2010}).

\bibitem{Paris2003}
\bibinfo{author}{Paris, M. G.~A.}, \bibinfo{author}{Cola, M.} \&
  \bibinfo{author}{Bonifacio, R.}
\newblock \bibinfo{title}{{Remote state preparation and teleportation in phase
  space}}.
\newblock \emph{\bibinfo{journal}{Journal of Optics B: Quantum and
  Semiclassical Optics}} \textbf{\bibinfo{volume}{5}},
  \bibinfo{pages}{S360--S364} (\bibinfo{year}{2003}).

\bibitem{Ye2013}
\bibinfo{author}{Ye, B.-L.}, \bibinfo{author}{Liu, Y.-M.}, \bibinfo{author}{Xu,
  C.-J.}, \bibinfo{author}{Liu, X.-S.} \& \bibinfo{author}{Zhang, Z.-J.}
\newblock \bibinfo{title}{{Quantum Correlations in a Family of Two-Qubit
  Separable States}}.
\newblock \emph{\bibinfo{journal}{Communications in Theoretical Physics}}
  \textbf{\bibinfo{volume}{60}}, \bibinfo{pages}{283--288}
  (\bibinfo{year}{2013}).

\bibitem{Girolami2011}
\bibinfo{author}{Girolami, D.}, \bibinfo{author}{Paternostro, M.} \&
  \bibinfo{author}{Adesso, G.}
\newblock \bibinfo{title}{{Faithful nonclassicality indicators and extremal
  quantum correlations in two-qubit states}}.
\newblock \emph{\bibinfo{journal}{Journal of Physics A: Mathematical and
  Theoretical}} \textbf{\bibinfo{volume}{44}}, \bibinfo{pages}{352002}
  (\bibinfo{year}{2011}).

\bibitem{Modi2010}
\bibinfo{author}{Modi, K.}, \bibinfo{author}{Paterek, T.},
  \bibinfo{author}{Son, W.}, \bibinfo{author}{Vedral, V.} \&
  \bibinfo{author}{Williamson, M.}
\newblock \bibinfo{title}{{Unified View of Quantum and Classical
  Correlations}}.
\newblock \emph{\bibinfo{journal}{Physical Review Letters}}
  \textbf{\bibinfo{volume}{104}}, \bibinfo{pages}{080501}
  (\bibinfo{year}{2010}).

\bibitem{Collett1987}
\bibinfo{author}{Collett, M.}, \bibinfo{author}{Loudon, R.} \&
  \bibinfo{author}{Gardiner, C.}
\newblock \bibinfo{title}{{Quantum Theory of Optical Homodyne and Heterodyne
  Detection}}.
\newblock \emph{\bibinfo{journal}{Journal of Modern Optics}}
  \textbf{\bibinfo{volume}{34}}, \bibinfo{pages}{881--902}
  (\bibinfo{year}{1987}).

\bibitem{Barnett2002}
\bibinfo{author}{Barnett, S.~M.} \& \bibinfo{author}{Radmore, P.~M.}
\newblock \emph{\bibinfo{title}{{Methods in Theoretical Quantum Optics}}}
  (\bibinfo{publisher}{Oxford Science Publications}, \bibinfo{year}{2002}).

\bibitem{Audretsch}
\bibinfo{author}{Audretsch, J.}
\newblock \emph{\bibinfo{title}{{Entangled Systems. New directions in quantum
  physics}}} (\bibinfo{publisher}{Wiley-VCH}, \bibinfo{address}{Germany},
  \bibinfo{year}{2007}), \bibinfo{edition}{first} edn.

\bibitem{Luo2008}
\bibinfo{author}{Luo, S.}
\newblock \bibinfo{title}{{Using measurement-induced disturbance to
  characterize correlations as classical or quantum}}.
\newblock \emph{\bibinfo{journal}{Physical Review A}}
  \textbf{\bibinfo{volume}{77}}, \bibinfo{pages}{022301}
  (\bibinfo{year}{2008}).

\bibitem{Nielsen00a}
\bibinfo{author}{Nielsen, M.} \& \bibinfo{author}{Chuang, I.~L.}
\newblock \emph{\bibinfo{title}{{Quantum Computation and Quantum Information}}}
  (\bibinfo{publisher}{Cambridge University Press}, \bibinfo{year}{2000}),
  \bibinfo{edition}{first} edn.

\bibitem{Olivares2013}
\bibinfo{author}{Olivares, S.}, \bibinfo{author}{Cialdi, S.},
  \bibinfo{author}{Castelli, F.} \& \bibinfo{author}{Paris, M. G.~a.}
\newblock \bibinfo{title}{{Homodyne detection as a near-optimum receiver for
  phase-shift-keyed binary communication in the presence of phase diffusion}}.
\newblock \emph{\bibinfo{journal}{Physical Review A - Atomic, Molecular, and
  Optical Physics}} \textbf{\bibinfo{volume}{87}}, \bibinfo{pages}{1--4}
  (\bibinfo{year}{2013}).

\bibitem{Kumar2012}
\bibinfo{author}{Kumar, R.} \emph{et~al.}
\newblock \bibinfo{title}{{Versatile wideband balanced detector for quantum
  optical homodyne tomography}}.
\newblock \emph{\bibinfo{journal}{Optics Communications}}
  \textbf{\bibinfo{volume}{285}}, \bibinfo{pages}{5259--5267}
  (\bibinfo{year}{2012}).

\bibitem{Ollivier2001}
\bibinfo{author}{Ollivier, H.} \& \bibinfo{author}{Zurek, W.~H.}
\newblock \bibinfo{title}{{Quantum Discord: A Measure of the Quantumness of
  Correlations}}.
\newblock \emph{\bibinfo{journal}{Physical Review Letters}}
  \textbf{\bibinfo{volume}{88}}, \bibinfo{pages}{017901}
  (\bibinfo{year}{2001}).

\bibitem{Henderson2001}
\bibinfo{author}{Henderson, L.} \& \bibinfo{author}{Vedral, V.}
\newblock \bibinfo{title}{{Classical, quantum and total correlations}}.
\newblock \emph{\bibinfo{journal}{Journal of Physics A: Mathematical and
  General}} \textbf{\bibinfo{volume}{34}}, \bibinfo{pages}{6899--6905}
  (\bibinfo{year}{2001}).

\bibitem{Giorda2012}
\bibinfo{author}{Giorda, P.}, \bibinfo{author}{Allegra, M.} \&
  \bibinfo{author}{Paris, M. G.~A.}
\newblock \bibinfo{title}{{Quantum discord for Gaussian states with
  non-Gaussian measurements}}.
\newblock \emph{\bibinfo{journal}{Physical Review A}}
  \textbf{\bibinfo{volume}{86}}, \bibinfo{pages}{052328}
  (\bibinfo{year}{2012}).

\bibitem{Campos1989}
\bibinfo{author}{Campos, R.}, \bibinfo{author}{Saleh, B.} \&
  \bibinfo{author}{Teich, M.}
\newblock \bibinfo{title}{{Quantum-mechanical lossless beam splitter: SU(2)
  symmetry and photon statistics}}.
\newblock \emph{\bibinfo{journal}{Physical Review A}}
  \textbf{\bibinfo{volume}{40}}, \bibinfo{pages}{1371--1384}
  (\bibinfo{year}{1989}).

\bibitem{leonhardt2003}
\bibinfo{author}{Leonhardt, U.}
\newblock \bibinfo{title}{{Quantum Physics of Simple Optical Instruments}}.
\newblock \emph{\bibinfo{journal}{Reports on Progress in Physics}}
  \textbf{\bibinfo{volume}{66}}, \bibinfo{pages}{1207} (\bibinfo{year}{2003}).

\bibitem{Leonhardt2010}
\bibinfo{author}{Leonhardt, U.}
\newblock \emph{\bibinfo{title}{{Essential Quantum Optics: From Quantum
  Measurements to Black Holes}}} (\bibinfo{publisher}{Cambridge University
  Press}, \bibinfo{address}{UK}, \bibinfo{year}{2010}), \bibinfo{edition}{first}
  edn.

\bibitem{Paris1996}
\bibinfo{author}{Paris, M. G.~a.}
\newblock \bibinfo{title}{{Displacement operator by beam splitter}}.
\newblock \emph{\bibinfo{journal}{Physics Letters, Section A: General, Atomic
  and Solid State Physics}} \textbf{\bibinfo{volume}{217}},
  \bibinfo{pages}{78--80} (\bibinfo{year}{1996}).

\bibitem{Ash65}
\bibinfo{author}{Ash, R.}
\newblock \emph{\bibinfo{title}{{Information Theory}}}
  (\bibinfo{publisher}{Interscience}, \bibinfo{address}{New York, USA},
  \bibinfo{year}{1965}), \bibinfo{edition}{1} edn.

\end{thebibliography}

\end{document}